\documentclass[useAMS,usenatbib]{mn2e}
\usepackage{graphicx}
\usepackage{amssymb}
\usepackage{lscape}

\newcommand{\apj}{ApJ}

\newcommand{\aap}{A\&A}

\newcommand{\apjs}{ApJS}
\newcommand{\MC}{\multicolumn}
\newcommand{\kms}{km\,s$^{-1}$}
\newcommand{\Te}{T$_{\rm e}$}
\newcommand{\sunn}{$_{\odot}$}
\newcounter{qub}
\newcommand{\qq}{\addtocounter{qub}{1}\arabic{qub}}

\DeclareRobustCommand{\ion}[2]{%
\relax\ifmmode
\ifx\testbx\f
{\mathrm{#1\,\textsc{#2}}}\else
{\mathrm{#1\,\mathsc{#2}}}\fi
\else\textup{#1\,{\mdseries\textsc{#2}}}%
\fi}

\title[Lynx-Cancer void galaxies. II. The element abundances]{Study of
galaxies in the Lynx-Cancer void. II. The element abundances}
\author[S. A. Pustilnik, A. L. Tepliakova, A. Y. Kniazev]
{S. A. Pustilnik,$^{1}$\thanks{sap@sao.ru (SAP), arina@sao.ru (ALT), akniazev@saao.ac.za (AYK)}
A. L. Tepliakova$^1$\footnotemark[1], A. Y. Kniazev$^{2}$ \\
$^1$ Special Astrophysical Observatory of RAS, Nizhnij Arkhyz,
Karachai-Circassia 369167, Russia\\
$^2$ South African Astronomical Observatory, Cape Town, SAR
}

\begin{document}

\label{firstpage}

\date{Accepted March 25, 2011. Received February 21, 2011 }

\pagerange{\pageref{firstpage}--\pageref{lastpage}} \pubyear{2011}

\maketitle

\begin{abstract}

In the framework of the study of the evolutionary status of
galaxies in the nearby Lynx-Cancer void, we present the results of the SAO
RAS 6-m telescope  spectroscopy for 20 objects in this region.
The principal faint line [OIII]$\lambda$4363~\AA, used to determine the
electron temperature and oxygen abundance (O/H) by the classical method,
is clearly detected in only  about 2/3 of the studied objects.
For the remaining galaxies this line is either faint or undetected.
To obtain the oxygen abundances in these galaxies we as well apply
the semi-empirical method by Izotov and Thuan, and/or the empirical
methods of Pilyugin et al., which are only employing the intensities
of sufficiently strong lines.
We also present our O/H measurements for 22 Lynx-Cancer
void galaxies, for which the suitable  Sloan Digital Sky Survey (SDSS)
spectra are available. In total, we present the combined O/H data
for 48 Lynx-Cancer void galaxies, including the data adopted from
the literature and our own earlier results.
We make a comparison of their locations on the (O/H)--M$_{\rm B}$
diagram with those of the dwarf galaxies of the Local Volume in
the regions with denser environment.  We infer that the majority
of galaxies from this void  on the average reveal an about 30\%
lower metallicity. In addition, a substantial fraction (not less
than 10\%) of the void dwarf galaxies have a much larger O/H
deficiency (up to a factor of 5). Most of them belong to the tiny
group of objects with the gas metallicity Z $<$Z\sunn/20 ~or~
12+$\log$(O/H)$\lesssim$7.35. The surface density of very
metal-poor galaxies \mbox{(Z $<$Z\sunn/10)} in this region of the
sky is 2--2.5 times higher than that, derived from the
emission-line galaxy samples in the Hamburg-SAO and the SDSS
surveys. We discuss possible implications of these results for the
galaxy evolution models.
\end{abstract}


\section{INTRODUCTION}
\label{sec:intro}

The relation between the properties of galaxies  and their environment is
studied and discussed for quite a long time.  Many aspects of the accelerated
galaxy evolution in the dense environment (galaxy clusters and compact groups)
are discovered and explained.  On the opposite edge of the galaxy density
range, for the most rarefied environment (or voids), a certain advance
was made in the study of galaxy properties, namely, their colour,
the rates of star formation, etc. mainly due to the huge data
supply by the Sloan Digital Sky Survey (SDSS) \cite{DR7} ~(see
\cite{PaperI} for details and references). However, these studies
did not deal directly with the evolutionary status of the void
galaxies. Since the effect of environment is expected to be the
largest for the galaxies of the smallest masses, it is reasonable
to study the galaxies in the nearby voids to trace any possible
differences in the evolution of dwarf galaxies. There one can make
a sample of the least massive and luminous objects. In Paper~I
\cite{PaperI} we described a sample of 79 galaxies in one of the
nearest voids, the Lynx-Cancer void, with total size in excess of
16~Mpc, and centre located 18~Mpc away. The sample consists mainly
of dwarf disc (irregular and late-type spiral) galaxies. About a
half of them belong to the Low Surface Brightness (LSB) galaxies.
This means that the central surface brightness of the underlying
disc of these objects, corrected for extinction and disc
inclination amounts to $\mu_{0,B,i,c} >$ $23^m/ \square ''$. The
absolute magnitude $M_{\rm B}$ of the sample galaxies lies between
$-11.9$ and $-18.4$, with the median value of about $-14.5$.
According to the preliminary estimates the sample is nearly
complete for the luminosities of \mbox{$-14 > M_{\rm B}$.}

The main goal of Paper~I  was to form a sufficiently deep and
large sample of galaxies in the individual void in order to study their
evolutionary parameters and spatial distribution. The evolutionary
status is characterized by the sufficiently easily assessed parameters:
the gas metallicity (in our case, the oxygen abundance (O/H) in the
 regions  of the current star formation (SF)) and the mass fraction of gas.
The latter parameter is measured from the direct
measurements of the gas mass via HI 21\,cm line flux (corrected for the
fraction of helium) and from the model-dependent stellar mass M$_{*}$.
The latter is determined from the optical
luminosity and the ratio M$_{*}$/L(opt), which depends on the galaxy colours.
One more parameter related to the galaxy evolution is the age of the oldest
population visible in the well-resolved galaxy images. It is usually assumed
that
the old stellar population is the most representative in the outer parts of
galaxies, outside the current SF regions. The latter mainly reside near the
galaxy centres or within the `optical' radius of the galaxy disc.
In most of galaxies, the population of the outer regions reveals colours
typical of stars  aged  \mbox{$T \gtrsim$ 10~Gyr}. Only in several galaxies
with a very low metallicity (I~Zw~18 and similar) this old population has not
been detected \cite{IT04,Guseva03}. Lately,  several galaxies with
younger populations, i.e. with $T_{\rm old}$ $\lesssim$ (1--3)~Gyr
have been discovered namely in the voids \cite{Pustilnik03,
SBS0335BTA, DDO68, DDO68_sdss, J0926}.

Within the ongoing project on the study of the evolutionary status
of the Lynx-Cancer void galaxies, we undertook the spectral survey of part
of our sample with the \mbox{6-m} BTA telescope of the
Special Astrophysical Observatory of Russian Academy of Sciences (SAO RAS)
with the goal to obtain the O/H of their gas.
For a part of the sample, we also used the SDSS spectra. For a comparative
analysis of the O/H parameter in the void galaxies with those
in a denser environment, we also included  several O/H estimates
for the void  galaxies published in the literature.

The amount of evidence for the existence of void galaxies,
representing a less evolved population was growing during the last decade
\cite{Peebles01,Pustilnik03,HS0837,DDO68,HS2134}.
However, this data concerned mainly the galaxies with active star formation
due to selection effects.
Therefore, we a need to address this issue using a more general approach,
which is mainly aimed on the analysis of more or less typical late-type
galaxies. As noted before, the less luminous and massive the galaxy,
the more it is susceptible to external perturbations.
Therefore, to study probable effects of the global environment,
one needs a sample of relatively nearby  galaxies.

The lay-out of the paper is as follows. Section~\ref{sec:obs} describes
the spectral observations and primary reduction of the BTA data,
the spectral data from the SDSS database and their analysis,
as well as the further analysis of emission-line spectra.
Section~\ref{sec:results} presents the description of the obtained results,
the tables of measured line intensities and derived physical parameters
and element abundances.
In Section~\ref{sec:dis} we  discuss all the results on the O/H for the
Lynx-Cancer
void galaxy sample accumulated so far, and compare them with O/H values for
the galaxies residing in denser regions.
The summary of our main results is given in Section~\ref{sec:summ}.
Appendix A   contains plots with the
spectra of all objects, obtained at the BTA or taken from the SDSS
database. Appendix B   presents the tables with relative line
intensities, as well as the derived O/H for all galaxies.

\section{Observations and reduction}
\label{sec:obs}

The observations presented in the paper were conducted at the BTA
in the period between  2002 and 2009 (see the journal of observations in
Table~\ref{Tab1_BTA}). Apart from the galaxies residing in the
Lynx-Cancer void, we also observed as back-up targets eight galaxies which
populate the adjacent denser regions.
They are presented  in the bottom of the same Table below the solid line.
The galaxy UGC~731 is the object  residing in the nearby Cepheus void.

The majority of observations were conducted with the multimode SCORPIO
instrument \citep{SCORPIO}, installed at the prime focus of the
BTA between January 12, 2007 and  February 19, 2009.
The grism VPHG550G and the 2K$\times$2K  EEV~42-40 CCD detector
were used for all observations except the nights  of January 21 and 22,
2009,  when the 2K$\times$4K EEV~42-90 CCD detector was used.
For several earlier observations we used the UAGS spectrograph,
installed in the prime focus of BTA, with the grating 400
grooves~mm$^{-1}$ and 1K$\times$1K pixel Photometrics-1024 CCD detector
\citep{Afanasiev95}.
The respective spectral ranges, spectral resolution, exposure times and
seeings for each object  are presented in Table~\ref{Tab1_BTA}.
The scale along the slit (after binning) was 0\farcs36 pixel$^{-1}$ in all
observations with the SCORPIO and \mbox{0\farcs40 pixel$^{-1}$} in the
observations with the UAGS. The object spectra were complemented by the
reference spectra of a \mbox{He--Ne--Ar} lamp for the wavelength calibration.
The spectral standard stars Feige~34, BD+28\degr4211, G191B2B and others from
the list of \citet{Bohlin96} were observed for the flux calibration several
times per night.
The slit was positioned on \ion{H}{ii} regions either found in the
literature \citep{vZee97,KK10} or identified in the images of the studied
galaxies, obtained with the SCORPIO with the
medium-width filter SED665 (FWHM=191~\AA, centred at $\lambda$6622~\AA).

For 22 Lynx-Cancer void galaxies (4 of them are from the BTA program)
we  derived useful element abundance estimates from the
spectra in the SDSS DR7 database.  The SDSS spectra were obtained through
a 3\arcsec\ round aperture with the multi-object fiber spectrograph,
and processed with the SDSS standard pipeline. A more detailed description
can be found in \citep{Gunn98,DR7}.

All spectral data reduction (for BTA observations)  was performed using the
technique, similar to that, described in  \citet{DDO68}.
The standard pipeline of long-slit spectra reduction uses the
IRAF\footnote{IRAF: the Image
Reduction and Analysis Facility is distributed by the National Optical
Astronomy Observatory, which is operated by the Association of Universities
for Research in Astronomy, Inc. (AURA) under cooperative agreement with the
National Science Foundation (NSF).}
and {\tt MIDAS}\footnote{MIDAS is an acronym for the European Southern
Observatory package -- Munich Image Data Analysis System. } codes.
It implies the following steps: the removal of cosmic ray hits,
bias subtraction,  flat-field correction, wavelength
calibration, night-sky background subtraction. Then, using the data on the
spectrophotometry of standard stars, we obtained the curves of
spectral sensitivity in a given night, and all spectra were transformed
to the absolute fluxes. Finally, individual 1D spectra of the target
H{\sc ii}-regions were
extracted  by the summation of several rows (typically 5--10,
 where the faint line [\ion{O}{iii}] $\lambda$4363, necessary to make
the temperature estimate $T_{\rm e}$  is visible). The emission
line intensities  with their errors were measured in the
individual 1D spectra using the method, described in detail in \citet{SHOC}.
Finally, the  element abundances were calculated using the procedures
described in \citet{Sgr} and references therein.

\begin{table*}
\begin{center}
\caption{Journal of the 6\,m telescope spectral observations}
\label{Tab1_BTA}
\begin{tabular}{llrccccccc} \\ \hline \hline
\MC{1}{c}{ Name }       &
\MC{1}{c}{ Date }       &
\MC{1}{c}{ Exp. }   &
\MC{1}{c}{ Wavelength } &
\MC{1}{c}{ Disp. } &
\MC{1}{c}{ Spec. } &
\MC{1}{c}{ Seeing }     &
\MC{1}{c}{ Air }     &
\MC{1}{c}{ Grism or}       &
\MC{1}{c}{ Detector }    \\

\MC{1}{c}{ }       &
\MC{1}{c}{ }       &
\MC{1}{c}{ time [s] }    &
\MC{1}{c}{ Range [\AA]  } &
\MC{1}{c}{ [\AA/px] } &
\MC{1}{c}{ res.(\AA) } &
\MC{1}{c}{ [$\prime\prime$] }    &
\MC{1}{c}{  mass    }    &
\MC{1}{c}{ grating  }    &
\MC{1}{c}{          }     \\

\MC{1}{c}{ (1) } &
\MC{1}{c}{ (2) } &
\MC{1}{c}{ (3) } &
\MC{1}{c}{ (4) } &
\MC{1}{c}{ (5) } &
\MC{1}{c}{ (6) } &
\MC{1}{c}{ (7) } &
\MC{1}{c}{ (8) } &
\MC{1}{c}{ (9) } &
\MC{1}{c}{ (10) } \\
\hline
\\[-0.3cm]
UGC~3475       & 2008.01.12  & 3$\times$900 & $ 3500-7500$ & 2.1 & 12.0 & 1.2 & 1.08 & VPHG550G  & 2K$\times$2K  \\
UGC~3476       & 2007.12.16  & 4$\times$900 & $ 3500-7500$ & 2.1 & 12.0 & 2.5 & 1.03 & VPHG550G  & 2K$\times$2K  \\
UGC~3501       & 2008.11.26  & 4$\times$900 & $ 3500-7500$ & 2.1 & 12.0 & 1.4 & 1.13 & VPHG550G  & 2K$\times$2K  \\
UGC~3600       & 2008.11.26  & 3$\times$900 & $ 3500-7500$ & 2.1 & 12.0 & 1.4 & 1.19 & VPHG550G  & 2K$\times$2K  \\
UGC~3672       & 2003.12.24  & 2$\times$900 & $ 3500-7500$ & 2.1 & 12.0 & 1.2 & 1.02 & G400      & 1K$\times$1K  \\
UGC~3698       & 2008.11.26  & 3$\times$900 & $ 3500-7500$ & 2.1 & 12.0 & 1.7 & 1.05 & VPHG550G  & 2K$\times$2K  \\
NGC~2337       & 2008.12.18  & 3$\times$420 & $ 3500-7500$ & 2.1 & 12.0 & 2.0 & 1.09 & VPHG550G  & 2K$\times$2K  \\
UGC~3817       & 2008.01.12  & 3$\times$900 & $ 3500-7500$ & 2.1 & 12.0 & 1.2 & 1.03 & VPHG550G  & 2K$\times$2K  \\
UGC~3860       & 2008.12.18  & 3$\times$600 & $ 3500-7500$ & 2.1 & 12.0 & 2.0 & 1.10 & VPHG550G  & 2K$\times$2K  \\
UGC~3876       & 2009.01.22  & 4$\times$900 & $ 3500-7500$ & 2.1 & 12.0 & 1.5 & 1.09 & VPHG550G  & 2K$\times$4K  \\
UGC~4117       & 2002.01.13  & 2$\times$900 & $ 3500-7500$ & 2.1 & 11.0 & 1.7 & 1.13 & G400      & 1K$\times$1K  \\
MCG7-17-19     & 2003.01.02  & 1$\times$1200& $ 3500-7500$ & 2.1 & 11.0 & 2.5 & 1.30 & G400      & 1K$\times$1K  \\
KUG~0821+321   & 2009.02.19  & 3$\times$900 & $ 3500-7500$ & 2.1 & 12.0 & 1.8 & 1.03 & VPHG550G  & 2K$\times$2K  \\
SDSS~J0843+4025& 2007.01.12  & 3$\times$900 & $ 3500-7500$ & 2.1 & 12.0 & 2.5 & 1.39 & VPHG550G  & 2K$\times$2K  \\
UGC~4704       & 2009.01.22  & 2$\times$900 & $ 3500-7500$ & 2.1 & 12.0 & 1.5 & 1.32 & VPHG550G  & 2K$\times$4K  \\
UGC~5272B      & 2008.01.12  & 4$\times$900 & $ 3500-7500$ & 2.1 & 12.0 & 1.0 & 1.12 & VPHG550G  & 2K$\times$2K  \\
UGC~5272       & 2002.01.12  & 1$\times$1200& $ 3500-7500$ & 2.1 & 11.0 & 1.6 & 1.04 & G400      & 1K$\times$1K  \\
SDSS~J1000+3032& 2009.01.21  & 3$\times$900 & $ 3500-7500$ & 2.1 & 12.0 & 1.4 & 1.39 & VPHG550G  & 2K$\times$4K  \\
UGC~5427       & 2007.01.12  & 5$\times$900 & $ 3500-7500$ & 2.1 & 12.0 & 2.4 & 1.07 & VPHG550G  & 2K$\times$2K  \\
UGC~5464       & 2008.01.12  & 3$\times$900 & $ 3500-7500$ & 2.1 & 12.0 & 1.0 & 1.29 & VPHG550G  & 2K$\times$2K  \\ \hline  
UGC~731        & 2008.11.26  & 4$\times$900 & $ 3500-7500$ & 2.1 & 12.0 & 1.5 & 1.02 & VPHG550G  & 2K$\times$2K  \\
SDSS~J0839+3140& 2009.01.22  & 2$\times$900 & $ 3500-7500$ & 2.1 & 12.0 & 1.5 & 1.18 & VPHG550G  & 2K$\times$4K  \\
UGC~4787       & 2009.01.22  & 4$\times$900 & $ 3500-7500$ & 2.1 & 12.0 & 1.9 & 1.06 & VPHG550G  & 2K$\times$4K  \\
KUG~1004+392   & 2008.12.18  & 3$\times$600 & $ 3500-7500$ & 2.1 & 12.0 & 1.5 & 1.02 & VPHG550G  & 2K$\times$2K  \\
UGC~5451       & 2008.01.13  & 3$\times$900 & $ 3500-7500$ & 2.1 & 12.0 & 1.2 & 1.05 & VPHG550G  & 2K$\times$2K  \\ 
SDSS~J1031+2801& 2009.02.19  & 3$\times$900 & $ 3500-7500$ & 2.1 & 12.0 & 1.7 & 1.29 & VPHG550G  & 2K$\times$2K  \\ 
NGC~3274       & 2009.02.19  & 2$\times$900 & $ 3500-7500$ & 2.1 & 12.0 & 1.7 & 1.60 & VPHG550G  & 2K$\times$2K  \\ 
UGC~5764       & 2009.01.22  & 3$\times$900 & $ 3500-7500$ & 2.1 & 12.0 & 2.0 & 1.22 & VPHG550G  & 2K$\times$4K  \\ 
UGC~6055       & 2009.02.19  & 2$\times$900 & $ 3500-7500$ & 2.1 & 12.0 & 1.7 & 1.41 & VPHG550G  & 2K$\times$2K  \\ 
\hline \hline \\[-0.2cm]
\end{tabular}
\end{center}
\end{table*}

\section{Results}
\label{sec:results}

\begin{figure}
 \centering
 \includegraphics[angle=-90,width=8cm]{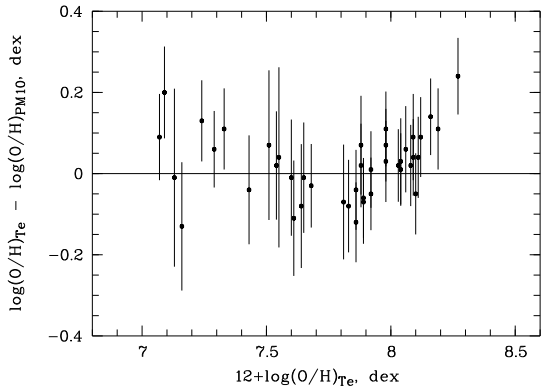}
  \caption{
\label{fig:testPM}
The illustration of the general consistency of O/H values derived
via the method of Pilyugin and Mattsson (PM10)  with
those  derived by the classic T$_{\rm e}$ method. Out of 42
galaxies, shown in the plot, 28 are the subsample of the nearby
non-BCG galaxies from the SHOC catalog. The data for 14 galaxies with
12+$\log$(O/H) $<$ 7.65, which are very few in the SHOC.
are adopted from other sources \citet{Kniazev03,IT07,vZee00,DDO68}.
}
\end{figure}

\begin{figure*}
 \centering
 \includegraphics[angle=-90,width=7cm]{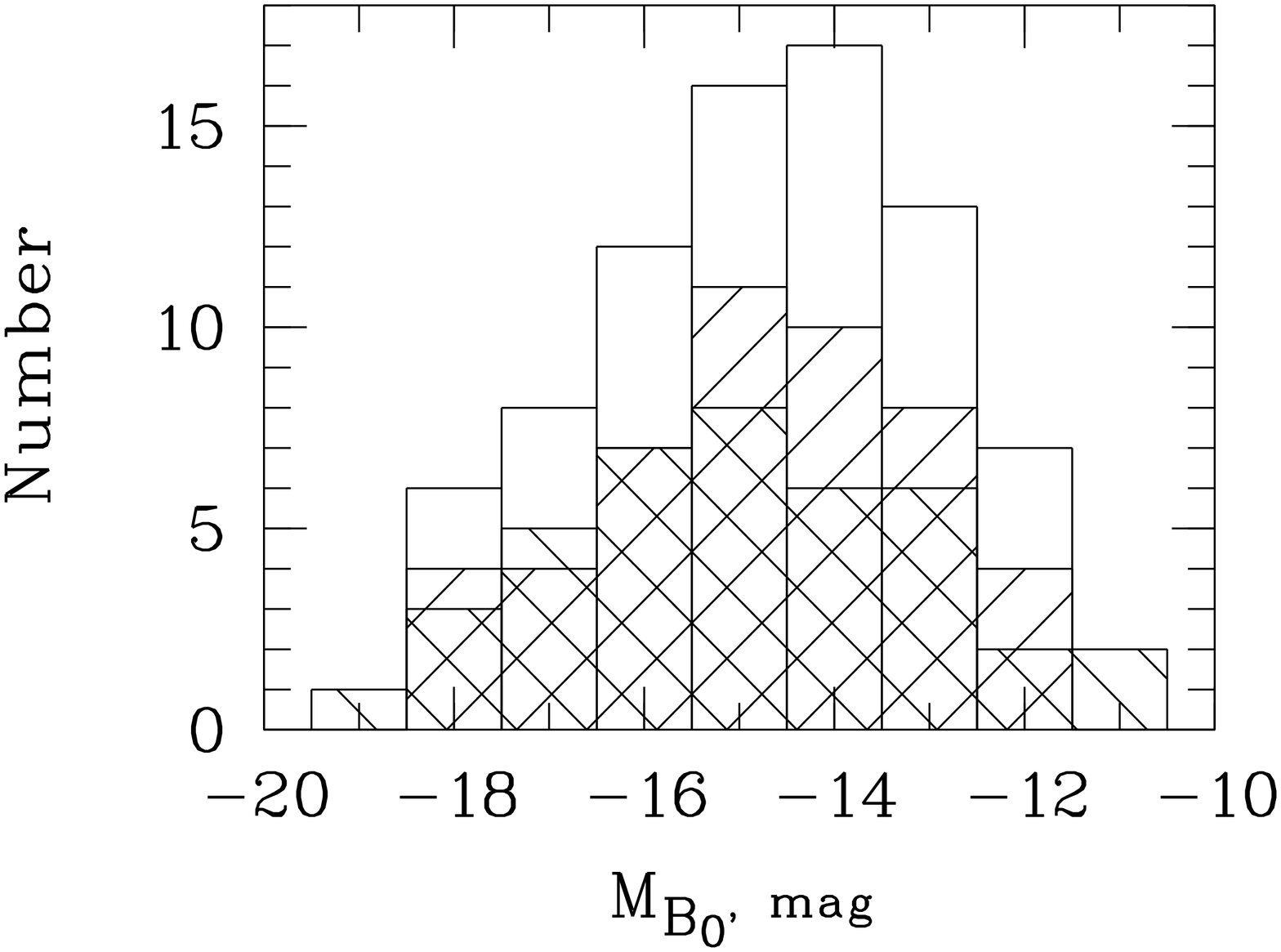}
 \includegraphics[angle=-90,width=7cm]{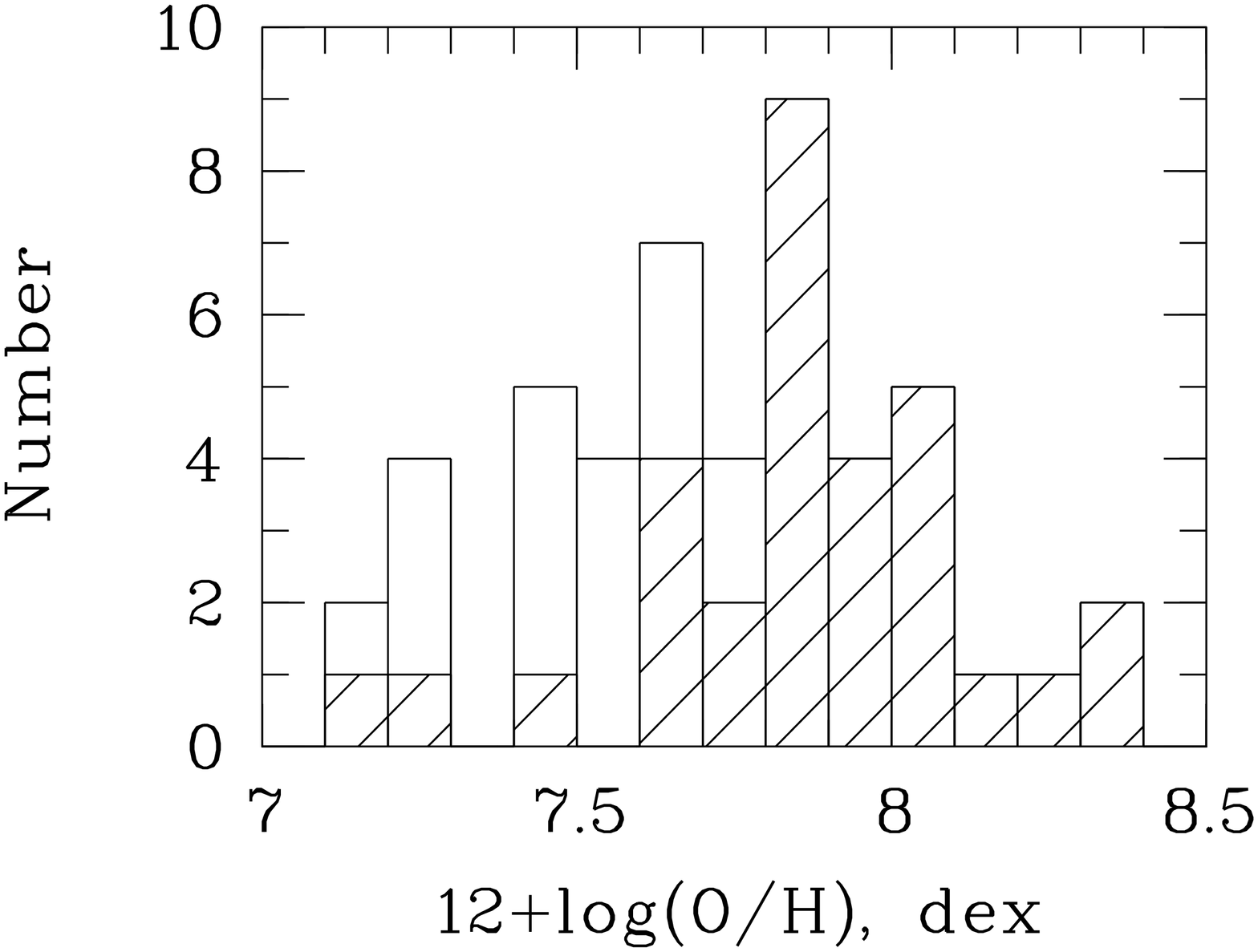}
  \caption{
\label{fig:histo}
{\bf Left panel:} the distribution of absolute magnitudes $M_{\rm B,0}$
for 48 galaxies
with known O/H (hatched) relative to the same distribution for all
79 Lynx-Cancer void galaxies. For a
comparison, the inverse hatching shows the distribution of $M_{\rm B,0}$
for 40 low-mass late-type galaxies from \citet{vZee06a}.
{\bf Right panel:} the distribution of all O/H values available to date
for 48 galaxies Lynx-Cancer void galaxies and for their subsample of 31
objects with M$_{\rm B} < -14.0$ (hatched), i.e. for the luminosity range,
in which  the void sample is expected to be almost complete.
}
\end{figure*}

\begin{figure*}
 \centering
 \includegraphics[angle=-90,width=14cm]{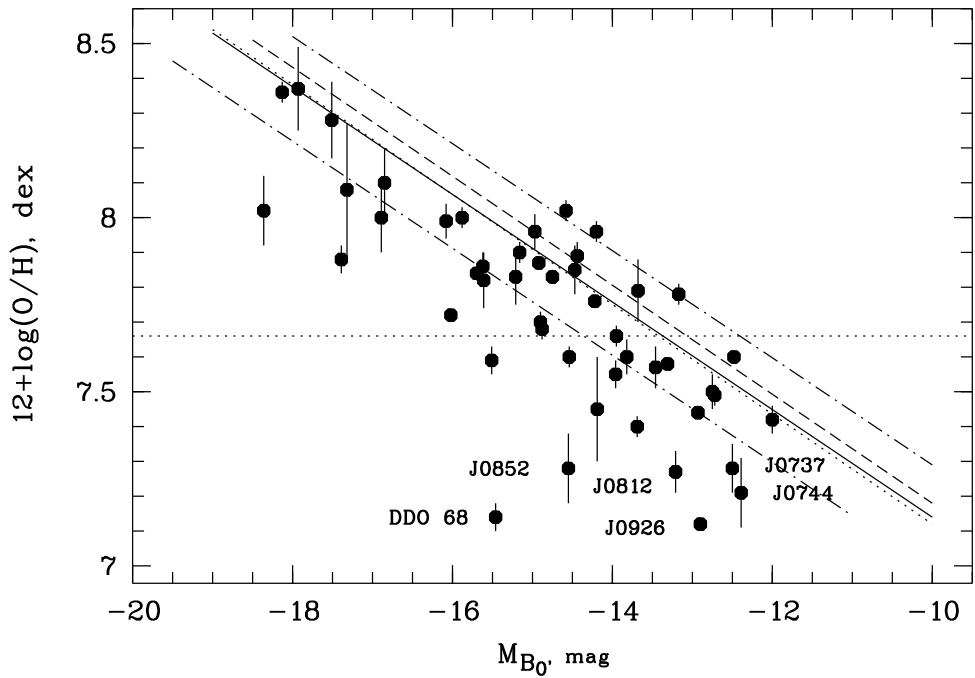}
  \caption{
\label{fig:Z_vs_M_BO}
The Luminosity-Metallicity (L--Z) relationship for 48 Lynx-Cancer void
galaxies.
The solid, dashed and dotted  lines mark the known linear fits for the L--Z
relationship, obtained fir 3 samples:  isolated late-type galaxies
\citet{vZee06a}, nearby dI galaxies \citet{vZee06b}, and nearby dI
and I galaxies from Lee et al. \citet{Lee03}.
Two dashed-dotted lines are drawn 0.15 dex lower and higher than
the solid line, for which the (computed in  \citet{vZee06a})
root mean squared deviation is 0.15 dex. If the O/H galaxy
distribution in this void corresponded to that, for which the
``standard'' L--Z relationship was obtained, one would expect that
about 1/6 of galaxies out of the total of 48 galaxies (i.e. 8
objects) would appear below the bottom dashed-dotted line. In the
reality though, 20 void galaxies made it into this region.}
\end{figure*}

The synopsis of the results of O/H measurements is laid out in
Table~\ref{tab:main}
and Figures~\ref{fig:histo},\ref{fig:Z_vs_M_BO} and discussed in
Section~\ref{sec:dis}. Below, we overview the obtained results in more detail.

Figures~A.1 and A.2 in Appendix\,A present the mosaic of 20
spectra of  20 Lynx-Cancer void
galaxies, observed at the BTA.
Tables~\,B.1--B.7 in Appendix\,B list  measured relative line
intensities of all relevant emission lines and their fluxes corrected
for extinction and the
underlying Balmer-line absorptions. This procedure is performed iteratively
by
the technique, described by \citet{Izotov94}. All the flux measurements follow
the method described by \citet{SHOC}.
The measured line flux values in the H$\beta$ line are given
in the units of \mbox{10$^{-16}$ erg\ s$^{-1}$cm$^{-2}$.}
Figure\,A.5
in Appendix\,A presents the BTA spectra of 10  H{\sc ii} regions
in 9 galaxies residing outside the Lynx-Cancer void. The relative
emission line fluxes for these galaxies, corrected for extinction
and the underlying Balmer-line absorptions are presented  in
Tables\,19.B--22.B of Appendix B.
Here under the name of each galaxy we indicated the coordinates of
the  H{\sc ii} region, for which the analyzed spectrum is
obtained.
Figures\,A.3 and A.4 of Appendix A 
present the mosaic of spectra of 20 Lynx-Cancer void galaxies,
extracted from the SDSS DR7 database.
In Tables\,8.B--14.B of Appendix B
  we present for the SDSS spectra the measured relative line
intensities of all relevant emission lines, and their fluxes
corrected for extinction and the underlying Balmer-line
absorptions, similar to those presented above for the BTA spectra.
The flux of the [O{\sc ii}]$\lambda$3737  line is shown in
parentheses, since this is not the measured value, but the one
recalculated from the line flux of the [O{\sc ii}]$\lambda$7320,
7330 lines, as described below.


Tables\,B.15 and B.16 in Appendix B
present  the electron temperatures  (based on the BTA spectra) we
measured in the emission zones of the O{\sc iii} and  O{\sc ii}
ions and their abundances along with the total O/H abundance,
derived for the line fluxes with the classic T$_{\rm e}$-method,
according to the scheme described \mbox{in \cite{Sgr}.} For the
cases when the [O{\sc iii}]$\lambda$4363 line is faint or
undetectable, we also apply the semi-empirical method of Izotov
and Thuan \cite{IT07} (based on the T$_{\rm e}$ estimate, derived
from the fluxes of the [O{\sc ii}]$\lambda$3727 and [O{\sc
iii}]$\lambda$4959, 5007 lines), which gives the O/H well
consistent with that derived by the classical \Te-method for the
range of \mbox{12+$\log$(O/H)$\lesssim$7.9.} However, according to
our tests, the semi-empirical method for the range of
\mbox{12+$\log$(O/H)$\gtrsim$7.9,} gives a systematically lower
O/H, with the difference of \mbox{$\Delta$(O/H)$\sim$--0.10~dex}
for the O/H, derived with the \Te-method in the range of
12+$\log$(O/H)$=7.9-8.1$. Therefore, when we used the
semi-empirical method for the above O/H range (as indicated, e.g.,
by other methods), the respective correction was applied. We also
show the O/H values derived via the empirical methods from
Pilyugin and Mattsson (hereinafter referred to as PM10)
\cite{PM10}, and Pilyugin et al. (hereinafter referred to as
PVT10) \cite{PVT10} for the cases when they have relatively small
errors. Similar data for the non-void BTA observations are
presented in Table\,B.23.



Tables\,B.17-B.18 of Appendix B present similar parameters for the
void galaxies, derived from the SDSS DR7 data. Since for the
galaxies with redshifts $z \lesssim$0.025 the line [O{\sc ii}]$\lambda$3727
is out of the range covered by the SDSS spectra,
one can use  the flux in the [O{\sc ii}]$\lambda$7320,7330  lines
to determine the O/H via the classical method
\cite{Kniazev03,SHOC}. For the cases when the [O{\sc
iii}]$\lambda$4363 line is faint or undetectable, the
semi-empirical method can be applied. In this case the line
intensities of [O{\sc ii}]$\lambda$7320, 7330 can be transformed
into the line intensity of   [O{\sc ii}]$\lambda$3727 through the
dependence between these parameters, presented in the graphic form
by Aller \cite{Aller84}. For the  low electron density ($N_{\rm e}
\lesssim 10^{2}$) the ratio of intensities of the [O{\sc ii}]$\lambda$3727
lines and the sum of intensities of the [O{\sc ii}]$\lambda$7320,7330 lines,
or the ``O-ratio''  as a function
of t for 1$<$t$<$2  (t$=$T$_{\rm e}$/10000) is well approximated
by the formula: {\small
$$\log(O-ratio)=1.77-2.06\times \log(t)+2.318\times \log(t)^2.$$}
The fluxes in the O{\sc ii}$\lambda$3727 line, listed in
Tables\,B.8-B.13 of Appendix\,B for the SDSS spectra are
extrapolated from fluxes in the  [O{\sc ii}]$\lambda$7320,7330
lines at the temperature T$_{\rm e}$,  adopted from a few
iterations. This is why they should be treated as approximate
estimates.


Apart from that, we used the empirical method for estimating the
O/H, recently proposed by Pilyugin and Mattsson (PM10)
\cite{PM10}, which exploits both the fluxes of strong lines
[O{\sc iii}]$\lambda$4959, 5007 and the fluxes of the
[N{\sc ii}]$\lambda$6548, 6584 and [S{\sc ii}]$\lambda$6716, 6730 lines.
To check how well the  method of Pilyugin and Mattsson \cite{PM10}
approximates the real O/H values in the studied galaxies, we made
a comparison of  O/H($T_{\rm e}$) values, known with the accuracy
of $\sigma_{O/H} \lesssim$0.06  with the O/H values derived via
the formulae \mbox{from \cite{PM10}.} To this end, we used 14
different measurements for the galaxies with
12+$\log$(O/H)$\lesssim$7.65 from the papers
\cite{Kniazev03, IT07, vZee00, DDO68} and 28 galaxies with the
best O/H measurements from the subsample of nearby non-BCG
galaxies from the SHOC catalog \cite{SHOC}, based on the SDSS
spectra. Figure\,1 illustrates the difference
\mbox{$\Delta~\log$(O/H)$ = \log$(O/H(PM10))--$\log$(O/H($T_{\rm e}$))}
versus
 O/H($T_{\rm e}$)  for these galaxies. In the
range of 12+$\log$(O/H) between 7.1 and 8.3, which fully covers
the range of O/H for our void sample, there is no systematic shift
between these two methods. The scatter around the zero line
corresponds to the estimated root mean square errors of both O/H
values.

\section{DISCUSSION}
\label{sec:dis}

Table~2 summarizes the information on the Lynx-Cancer void
galaxies with the currently available O/H data, which can be used
for the subsequent statistical analysis. The following information
is presented. Column 1 gives the galaxy name, 2 and 3---right
ascension and declination (J2000); columns 4 and 5 list the J2000
coordinates of the  H{\sc ii} region, for which the spectrum was
obtained; column 6 gives the heliocentric velocity (in
km~s$^{-1}$); column 7---the distance in Mpc relative to the Local
Group centre. Columns 8 and 9 present the adopted total
$B$-magnitude  and the respective absolute magnitude. All these
parameters (except for the coordinates of the H{\sc ii} region)
are taken from Table~2 of Paper~I. In cases when the O/H was
adopted from literature, we do not list the region coordinates and
refer the reader to the original paper. In Column 10 we present
the value of O/H adopted for the further analysis. It can come
from different sources, indicated in Column 11 as follows: (1) the
weighted mean O/H for two or more H{\sc ii} regions in the same
galaxy from the BTA data, evaluated with the classical T$_{\rm
e}$-method; (2) the most probable O/H  estimate from the BTA
spectra, based on the combination of results of the classical
T$_{\rm e}$-method, the semi-empirical method \cite{IT07} and the
empirical methods \mbox{from \cite{PM10, PVT10};} (3) the same as
in (2), but based on the SDSS spectra; \mbox{(4)} O/H, adopted
from the literature and rescaled to the new scale \cite{Izotov06}
(if necessary).

The left plot of Fig.\,2 presents the distributions of 48
Lynx-Cancer void galaxies with known O/H (the hatched histogram)
compared to the statistics of the full void sample of 79 galaxies,
which demonstrates the galaxy fractions with known O/H in each
luminosity bin. For a comparison, the histogram with the inverse
hatching\footnote{Right hatching---from the bottom left corner to
the top right corner, the inverse hatching is perpendicular to
it.} and in a wider range shows a similar distribution for 40
sample galaxies from \citet{vZee06a}, used to check the effect of
the neighbourhood (see below). The void galaxies with known O/H are
distributed more or less homogeneously over the whole luminosity
range, with the average fraction of about 0.6. This allows to
conclude that the real O/H distribution of the void galaxies does
not differ significantly from that shown in the right panel for 48
galaxies with known O/H. In this panel the hatched histogram
demonstrates the O/H distribution for 31 galaxies with M$_{\rm B} <$--14.0,
i.e. for the range, where the void galaxy sample is
expected to be almost complete. Therefore, the tail of
distribution with  very metal-poor galaxies (7.1 $< 12+\log$(O/H) $<$ 7.5)
should be treated as a trustworthy.

\subsection{L--Z relationship for void galaxies}

To examine how much the metallicities of the Lynx-Cancer void galaxies differ
from those of the galaxies, residing in denser environments, one can use
the empirical relationships between the O/H and M$_{\rm B}$ for the dwarf
galaxy
samples in the Local Volume and its surroundings \citep{vZee06a}.
However, we need to eliminate the systematic  differences in the O/H scale,
i.e. to account for
small changes in the O/H scale used in recent papers (including the O/H
values, presented here) with respect to the scale used
in earlier papers, on which the ``standard'' relationship O/H--$M_{\rm B}$
we further use is based.
This O/H scale variation, suggested by \citet{Izotov06}, is related to
the changes in the atomic constants, used in the calculations.
When the respective correction is made, the original
formula for the L--Z relationship from \citet{vZee06a}
$$ 12+\log(O/H)=5.65-0.149 \cdot M_{\rm B} $$
transforms to
$$ 12+\log(O/H)=5.60--0.154 \cdot M_{\rm B} $$
\noindent This relationship is illustrated in Fig.~3 
by a straight solid line.

This figure also demonstrates similar relationships derived for
different samples of late-type galaxies within the Local Volume.
The dashed line, running a bit higher than the solid line
describes the relationship for the sample from  \cite{vZee06b}.
The dotted line, running almost along the solid line shows a
similar relationship derived by Lee et al. \cite{Lee03} for the
late-type dwarf galaxies from the nearby groups within the radius
of 5~Mpc. Since  all the three lines are very close to each other,
they probably describe the same general relationship. For the
further comparison with the ``standard'' relationship we adopt the
one, derived in \cite{vZee06a}, since it has the smallest scatter,
\mbox{$\sigma$(O/H)$\sim$0.15~dex.}  To demonstrate a significant
deviation of void galaxies from the  ``standard'' relationship, we
also plotted in Fig.~3 the dash-dotted lines, running 0.15~dex
lower and higher than the solid ``standard'' line \mbox{from
\cite{vZee06a}.}

The analysis of the O/H values of  void galaxy in the L--Z diagram shows
a clear systematic shift to the lower median values relative to the
`standard' line for every bin of the M$_{\rm B}$ range.
Even if we exclude from the consideration
 six objects with the lowest O/H (i.e. with 12+$\log$(O/H) $\lesssim$7.35)
as probably having atypical evolution
histories or relatively small ages, the remaining galaxies will be mostly
distributed below  the `standard' line. Their distribution becomes more
symmetrical relative to the line parallel to the `standard' L--Z line,
shifted  down by about 0.12~dex. This implies that about
85--90\%  quite typical galaxies, populating this void show a sizable
deficiency of oxygen (on the average not less \mbox{than 30\%})
relative to the galaxies used in constructing the `standard'
L--Z relation in the Local Volume. An alternative
interpretation of the observed shift as a luminosity increase at
the same metallicity seems unlikely. Indeed, an average shift by
about 0.12~dex at a fixed $M_{\rm B}$ corresponds to a shift of
0.75~mag in $M_{\rm B}$ at the fixed O/H value, i.e. to the
brightening by a factor of 2. This is a typical value for BCG-type
galaxies (see, e.g., \cite{Papa96}). However, in the discussed
sample of void galaxies only two galaxies belong to this type:
HS~0822+3542 and HS~1013+3809. The position of the former can be
explained by a shift with respect to the ``standard'' line (via
brightening) by about 1~magnitude. However, for the latter we
would have to assume a brightening by about 2.5~magnitudes. The
rest of the galaxies, as described in Paper~I, belong to the late
spirals and irregulars, in which the luminosities of the brightest
H{\sc ii} regions are small compared to the total luminosity of
the galaxy.

\headsep 1cm
\renewcommand{\baselinestretch}{0.9}

\setcounter{qub}{0}

\begin{table*}

\begin{center}

\caption{\label{tab:main}
Main galaxy parameters in the Lynx-Cancer void with the known O/H values}
\vspace{0.3cm}

\footnotesize{
\begin{tabular}{r|lc|cc|c|r|r|r|r|cl}
\hline 
\multicolumn{1}{c|}{No.}               & \multicolumn{1}{c|}{Name
or} & \multicolumn{2}{c|}{Coordinates (J2000)} &
\multicolumn{2}{c|}{Coordinates (J2000)} &
\multicolumn{1}{c|}{$V_{\rm hel}$,} & \multicolumn{1}{c|}{D,} &
\multicolumn{1}{c|}{B$_{\rm tot}$,}
& \multicolumn{1}{c|}{M$_{\rm B}^{\dagger}$,}  &
\multicolumn{1}{r|}{12+$\log$(O/H)}  &
\multicolumn{1}{l}{Source}       \\
& \multicolumn{1}{c|}{prefix}  & \multicolumn{2}{c|}{of galaxy}  &
\multicolumn{2}{c|}{of H{\sc ii} region}  &
\multicolumn{1}{r|}{km/s} & \multicolumn{1}{r|}{Mpc}  &
\multicolumn{1}{r|}{mag}  & \multicolumn{1}{r|}{mag}  &
\multicolumn{1}{c|}{adopt.}  &
\multicolumn{1}{l}{O/H }  \\
& \multicolumn{1}{c|}{ (1) }  & \multicolumn{1}{c}{ (2) }  &
\multicolumn{1}{c|}{ (3) }  & \multicolumn{1}{c}{ (4) }  &
\multicolumn{1}{c|}{ (5) }  & \multicolumn{1}{c|}{ (6) }  &
\multicolumn{1}{c|}{ (7) }  & \multicolumn{1}{c|}{ (8) }  &
\multicolumn{1}{c|}{ (9) }  & \multicolumn{1}{c|}{ (10) }  &
\multicolumn{1}{c}{ (11) }  \\
\hline
\hline
\qq&  UGC3475       & 06 30 28.86&  +39 30 13.6& 06 30 27.02&  +39 30 26.3& 487&   10.30& 14.97&  --15.88  & 8.00$\pm$0.03 &  2        \\ 
\qq&  UGC3476       & 06 30 29.22&  +33 18 07.2& 06 30 29.27&  +33 18 05.0& 469&    9.84& 14.96&  --16.02  & 7.72$\pm$0.02 &  2        \\ 
\qq&  UGC3501       & 06 38 38.40&  +49 15 30.0& 06 38 40.24&  +49 15 28.6& 449&   10.07& 17.20&  --13.31  & 7.58$\pm$0.02 &  2        \\ 
\qq&  UGC3600       & 06 55 40.00&  +39 05 42.8& 06 55 41.90&  +39 06 21.7& 412&    9.30& 16.18&  --13.95  & 7.66$\pm$0.03 &  2        \\ 
\qq&  UGC3672       & 07 06 27.56&  +30 19 19.4& 07 06 25.46&  +30 19 36.4& 994&   16.93& 15.40&  --16.08  & 7.99$\pm$0.05 &  2        \\ 
\qq&  UGC3698       & 07 09 16.80&  +44 22 48.0& 07 09 18.15&  +44 22 43.7& 422&    9.60& 15.41&  --14.92  & 7.87$\pm$0.02 &  2        \\ 
\qq&  NGC2337       & 07 10 13.60&  +44 27 25.0& 07 10 12.29&  +44 27 43.7& 436&    9.79& 13.48&  --16.85  & 8.10$\pm$0.10 &  1        \\ 
\qq&  UGC3817       & 07 22 44.48&  +45 06 30.7& 07 22 44.84&  +45 06 41.8& 437&    9.82& 15.96&  --14.44  & 7.89$\pm$0.04 &  2        \\ 
\qq&  SDSS          & 07 23 01.42&  +36 21 17.1&            &             & 885&   14.38& 17.01&  --14.19  & 7.45$\pm$0.15 &  4d       \\ 
\qq&  UGC3860       & 07 28 17.20&  +40 46 13.0& 07 28 19.44&  +40 46 29.9& 354&    7.81& 14.96&  --14.75  & 7.83$\pm$0.02 &  2        \\ 
\qq&  UGC3876       & 07 29 17.49&  +27 54 01.9& 07 29 17.36&  +27 53 16.1& 854&   15.01& 13.70&  --17.39  & 7.88$\pm$0.04 &  2        \\ 
\qq&  SDSS          & 07 30 58.90&  +41 09 59.8& 07 30 58.90&  +41 09 59.8& 874&   15.70& 16.67&  --14.58  & 8.02$\pm$0.03 &  3        \\ 
\qq&  SDSS          & 07 37 28.47&  +47 24 32.8&            &             & 476&   10.42& 18.06&  --12.50  & 7.28$\pm$0.07 &  4d       \\ 
\qq&  SDSS          & 07 44 43.72&  +25 08 26.6& 07 44 43.72&  +25 08 26.6& 749&   12.95& 18.35&  --12.39  & 7.21$\pm$0.10 &  3        \\ 
\qq&MCG9-13-56      & 07 47 32.10&  +51 11 29.0& 07 47 33.18&  +51 11 24.7& 439&   10.00& 15.48&  --14.90  & 7.70$\pm$0.03 &  3,4b     \\ 
\qq&  UGC4117       & 07 57 25.98&  +35 56 21.0& 07 57 26.31&  +35 56 19.2& 773&   14.12& 15.34&  --15.61  & 7.82$\pm$0.08 &  2        \\ 
\qq&  UGC4148       & 08 00 23.68&  +42 11 37.0& 08 00 23.68&  +42 11 37.0& 716&   13.55& 15.63&  --15.21  & 7.83$\pm$0.08 &  3        \\ 
\qq& NGC2500        & 08 01 53.30&  +50 44 15.4& 08 01 55.38&  +50 44 38.5& 504&   10.88& 12.23&  --18.13  & 8.36$\pm$0.03 &  3        \\ 
\qq& MCG7-17-19     & 08 09 36.10&  +41 35 40.0& 08 09 37.64&  +41 35 29.5& 704&   13.37& 16.65&  --14.20  & 7.96$\pm$0.02 &  2,3      \\ 
\qq& SDSS           & 08 10 30.65&  +18 37 04.1& 08 10 30.65&  +18 37 04.1&1483&   23.05& 18.29&  --13.68  & 7.79$\pm$0.09 &  3        \\ 
\qq& SDSS           & 08 12 39.53&  +48 36 45.4& 08 12 39.53&  +48 36 45.4& 521&   11.05& 17.23&  --13.21  & 7.27$\pm$0.06 &  4b,3     \\ 
\qq& NGC2537        & 08 13 14.73&  +45 59 26.3& 08 13 13.05&  +45 59 26.3& 445&    9.86& 12.27&  --17.93  & 8.37$\pm$0.12 &  3        \\ 
\qq& IC2233         & 08 13 58.93&  +45 44 34.3&            &             & 553&   10.70& 13.05&  --17.32  & 8.08$\pm$0.19 &  4a       \\ %
\qq& NGC2541        & 08 14 40.18&  +49 03 42.1& 08 14 47.53&  +49 04 00.0& 548&   12.00& 12.25&  --18.36  & 8.02$\pm$0.10 &  3,4a     \\ 
\qq& NGC2552        & 08 19 20.14&  +50 00 25.2&            &             & 524&   11.11& 12.92&  --17.51  & 8.28$\pm$0.11 &  4a       \\ 
\qq& KUG~0821+321   & 08 25 04.90&  +32 01 05.1& 08 25 04.90&  +32 01 05.1& 648&   12.25& 16.10&  --14.54  & 7.60$\pm$0.03 &  2        \\ 
\qq& HS~0822+3542   & 08 25 55.43&  +35 32 31.9&            &             & 720&   13.49& 17.92&  --12.93  & 7.44$\pm$0.02 & 4e        \\ 
\qq&  SDSS          & 08 43 37.98&  +40 25 47.2& 08 43 37.98&  +40 25 47.2& 614&   12.05&17.83&   --12.72  & 7.49$\pm$0.03 &  2,3,4b   \\ 
\qq& SDSS           & 08 52 33.75&  +13 50 28.3&            &             &1511&   23.08&17.43&   --14.55  & 7.28$\pm$0.10 &  4d       \\ 
\qq&  UGC4704       & 08 59 00.28&  +39 12 35.7& 08 59 00.28&  +39 12 35.7& 596&   11.74&15.51&   --14.97  & 7.96$\pm$0.05 &  2        \\ 
\qq&  SDSS          & 08 59 46.93&  +39 23 05.6&            &             & 588&   11.63&16.98&   --13.46  & 7.57$\pm$0.06 &  4b       \\ 
\qq&  SDSS          & 09 11 59.43&  +31 35 35.9&            &             & 750&   13.52&17.97&   --12.75  & 7.50$\pm$0.05 &  4b       \\ 
\qq&  SDSS          & 09 26 09.45&  +33 43 04.1&            &             & 536&   10.63&17.34&   --12.90  & 7.12$\pm$0.02 &  4f       \\ 
\qq& SDSS           & 09 28 59.06&  +28 45 28.5& 09 28 59.06&  +28 45 28.5&1229&   19.90&16.70&   --14.88  & 7.68$\pm$0.03 &  3        \\ 
\qq&  SDSS          & 09 31 36.15&  +27 17 46.6& 09 31 36.15&  +27 17 46.6&1505&   23.60&17.98&   --13.96  & 7.55$\pm$0.04 &  3        \\ 
\qq&  SDSS          & 09 40 03.27&  +44 59 31.7& 09 40 03.27&  +44 59 31.7&1246&   20.71&17.96&   --13.69  & 7.40$\pm$0.03 &  3        \\ 
\qq&  KISSB23       & 09 40 12.67&  +29 35 29.3& 09 40 12.67&  +29 35 29.3& 505&   10.21&16.32&   --13.82  & 7.60$\pm$0.05 &  3,4b     \\ 
\qq& SDSS           & 09 47 18.35&  +41 38 16.4& 09 47 18.35&  +41 38 16.4&1389&   22.56&17.61&   --14.22  & 7.76$\pm$0.02 &  3        \\ 
\qq&  UGC5272B      & 09 50 19.49&  +31 27 22.3& 09 50 19.49&  +31 27 22.3& 539&   10.27&17.68&   --12.48  & 7.60$\pm$0.02 &  2,3      \\ 
\qq&  UGC5272       & 09 50 22.40&  +31 29 16.0& 09 50 21.35&  +31 29 16.6& 520&   10.30&14.46&   --15.70  & 7.84$\pm$0.02 &  2,3      \\ 
\qq&  SDSS          & 09 51 41.67&  +38 42 07.3& 09 51 41.67&  +38 42 07.3&1435&   23.07&17.42&   --14.47  & 7.85$\pm$0.07 &   3       \\ 
\qq&  UGC5340       & 09 56 45.70&  +28 49 35.0&            &             & 502&    9.86&14.60&   --15.45  & 7.14$\pm$0.03 &   4b      \\ 
\qq&  SDSS          & 10 00 36.50&  +30 32 09.8& 10 00 36.50&  +30 32 09.8& 501&    9.90&18.06&   --12.00  & 7.42$\pm$0.04 &  2,3      \\ 
\qq&  UGC5427       & 10 04 41.05&  +29 21 55.2& 10 04 40.44&  +29 21 35.9& 498&    9.79&14.89&   --15.16  & 7.90$\pm$0.03 &  2        \\ 
\qq&  UGC5464       & 10 08 07.70&  +29 32 34.4& 10 08 07.70&  +29 32 34.4&1003&   16.90&15.62&   --15.62  & 7.86$\pm$0.04 &  2        \\ 
\qq&  SDSS          & 10 10 14.96&  +46 17 44.1& 10 10 14.96&  +46 17 44.1&1092&   18.58&18.20&   --13.17  & 7.78$\pm$0.03 &   3       \\ 
\qq&  UGC5540       & 10 16 21.70&  +37 46 48.7& 10 16 21.70&  +37 46 48.7&1162&   19.16&14.60&   --16.89  & 8.00$\pm$0.10 &  3        \\ 
\qq&  HS 1013+3809  & 10 16 24.50&  +37 54 46.0&            &             &1173&   19.30&15.99&   --15.51  & 7.59$\pm$0.04 &  4g       \\ 
  \hline
\multicolumn{11}{l}{$^{\dagger}$ M$_{\rm B}$ from [2]. References for O/H: 1, 2, 3, 4---see table's description.  References in parenthesis:~~~a---\cite{SHOC}; }  \\
\multicolumn{11}{l}{~~~~~b---\cite{IT07}; c---\cite{vZee97}; d---\cite{LSB_void}; e---\cite{Pustilnik03}; }  \\
\multicolumn{11}{l}{~~~~~f---\cite{J0926}; g---\cite{Kniazev99}.}
\end{tabular}
}
\end{center}
\end{table*}


A large scatter of actively star-forming galaxies in the L--Z diagram,
especially in the region of very low metallicities
(12+$\log$(O/H)$<$7.65), was already noted in the literature
\citep[e.g., ][]{Kniazev03,Guseva09}. For such galaxies, several
possible evolutionary scenarios were suggested. One of them implies the
genuine non-evolved galaxies. It is likely that namely in the void
environments with low density of the surrounding galaxies,
such ``young'' very metal-poor galaxies could greatly contribute to the
general galaxy population and hence lead to the greater scatter of the galaxy
localisation in L--Z diagram.

\subsection{The most metal-poor Lynx-Cancer void galaxies}

Coming back to the subject of the most metal-deficient and unusual dwarf
galaxies, let us discuss them in more detail.
These include a Blue Compact Galaxy  HS 0822+3542 with the lowest metallicity
(12+$\log$(O/H)=7.44) among all the BCG galaxies of the Local Volume and
its surroundings, and its companion, a very blue (and presumably relatively
young) galaxy of low surface brightness (LSBD) SAO 0822+3545
\citep[][and a paper in preparation]{Pustilnik03}.
Other representatives of this void galaxy population are DDO~68 and SDSS
J0926+3343 with 12+$\log$(O/H), amounting to 7.14 and 7.12, respectively.
These are the most metal-poor galaxies known after
SBS~0335--052W and SDSS~J0015+0104 \citep{Izotov09, Guseva09}.
Situated on the mutual distance of only 1.6 Mpc, both of them are
also unusual by the  blue colours
of their outer parts, indicating the ages of the oldest visible stellar
population of $\lesssim$1--3~Gyr \citep{DDO68_sdss,J0926}, what is in
drastic contrast with the situation in the absolute majority of other dwarf
galaxies known to date.

In addition to these galaxies, the following void objects are among the most
metal-poor: SDSS J0812+4836, with O/H=7.27 dex \citep{IT07}, two LSBD galaxies
from our recent paper \citet{LSB_void}  SDSS
J0737+4724 and SDSS J0852+1350 both with 12+$\log$(O/H)=7.28\,dex.
Finally, for two more Lynx-Cancer void dwarf galaxies, --
SDSS J0744+2508 and J0940+4459, we have obtained now a very low oxygen
abundance estimate of \mbox{12+$\log$(O/H)$\sim$7.2--7.4.} .
Therefore, about a half-dozen of all the 48 void galaxies with known
O/H belong to the range of the lowest gas metallicities
(12+$\log$(O/H)$\lesssim$7.3). Out of many thousands of late-type
galaxies with known O/H, derived with different methods, only a
few such galaxies have been found by now (I~Zw~18, SBS~0335--052E
and W, UGCA~292).

It is worth to compare the surface density of very metal-poor (or eXtremely
Metal-Deficient, XMD -- \mbox{12+$\log$(O/H)$\leq$7.65)} galaxies
found in the Lynx-Cancer void sample with the estimates presented by us
for two different samples of emission-line galaxies (ELGs). The first estimate
was done for the ELG sample from the Hamburg-SAO Survey (HSS)
\citep[][ and references therein]{HSS1,HSS6}.
Most of these ELGs were selected according to criteria:
$B_{\rm tot} <$18.5 and
EW([O{\sc iii}]$\lambda$5007)$>$50~\AA. The latter roughly corresponds to
the criterion EW(H$\beta$)$\gtrsim$10~\AA. \citet{Pustilnik_ASS} have found
that the density of XMD galaxies in this sample is about 4 per
\mbox{1000\,$\square \degr$.}
A similar density estimate was derived by \citet{Kniazev03},
based on the galaxy sample selected from the SDSS (roughly similar to the HSS
by  the limiting $B$-magnitude) based on the strength of the H$\beta$
emission line (EW(H$\beta$)$>$10~\AA). Both these estimates were based on
the XMD galaxies, for which the O/H was derived via the classic
\mbox{T$_{\rm e}$-method.}

The region of the celestial sphere, onto which the Lynx-Cancer
void in projected contains 19 XMD galaxies, inhabiting this void.
Five more XMD galaxies fall in this sky
region from both background and foreground. This corresponds to the total
surface density of such galaxies of about 10 per 1000\,$\square \degr$,
which is almost a factor of 2.5 larger than the estimates based on
the ELG galaxy samples. This can be caused by at least two factors, or
a combination therein. The first factor is omission of the selection
criterion by rather strong H$\beta$ emission. The use of semi-empirical and
empirical methods
allows one to get more or less reliable estimates of O/H for the cases of
relatively weak H$\beta$ emission. The second reason is probably directly
linked with the fact that a nearby void, populated by a large fraction
of dwarf galaxies with slow evolution, is projected onto the given sky region.

The full volume of the Lynx-Cancer void represents only a small fraction
(about 5\%) of the sphere with $R=26$\,Mpc, which comprises this void.
The existence of such an unusually strong  concentration of the most
metal-poor galaxies in a small cell of nearby Universe is indicative of
the physical relation
between the evolutionary status of low-mass late-type galaxies and the type
of their global environment.
The discovery of several unusual dwarfs with relatively small ages of the
oldest
visible stellar population among the Lynx-Cancer void objects
gives additional evidences for such a relation.

\section{Summary}
\label{sec:summ}

1) Based on the results of BTA spectroscopy we present the oxygen abundances
for twenty galaxies residing in the nearby Lynx-Cancer void and for nine
galaxies situated outside this void.  For 14 galaxies, the abundances of
oxygen are derived via the classic T$_{\rm e}$-method with the accuracy of
O/H varying in the range of 0.06 to 0.15\,dex. For the remaining H{\sc ii}
regions with a faint or undetected [O{\sc iii}]$\lambda$4363 line, the
abundance of oxygen is estimated via the semi-empirical and empirical methods.

2) For 4 Lynx-Cancer void galaxies with a relatively strong
[O{\sc iii}]$\lambda$4363 line we use the spectra from the SDSS DR7
database to derive the oxygen abundances via the modified classic
 T$_{\rm e}$ method (from the  lines of [O{\sc ii}]$\lambda\lambda$7320,7330).
For the remaining 6 galaxies with a faint or undetected
[O{\sc iii}]$\lambda$4363 line, we derive O/H via the empirical
methods. We cross-compare all the O/H estimates derived from different
measurements (BTA, SDSS and from literature) and via various methods, and
calculate the weighted mean value to adopt the most reliable estimate of
the galaxy O/H.

3) Combining the data for all 48 galaxies in the Lynx-Cancer void with
available
O/H and M$_{\rm B}$ (including about a dozen of objects from the literature
and our other papers), we analyze their locations on the O/H--M$_{\rm B}$
diagram. A comparison of these data with the linear relationship between
$\log$(O/H) and $M_{\rm B}$, derived from three samples of low-mass late-type
galaxies residing in the Local Volume and its immediate
neighbourhood (the ``standard'' L--Z relationship) leads to the
following conclusions. The first: around 12\% of the studied void
galaxies (namely, the objects with 12+$\log$(O/H)$\lesssim$7.3)
demonstrate a large oxygen deficiency relative to the abundances,
expected from their luminosities in the ``standard'' L--Z
relationship (up to a few times). The second: the majority of the
remaining (approximately 88\%) void galaxies possess the metal
abundances systematically lower than   expected from the
``standard'' L--Z ratio, with the average O/H shift of the group
amounting to about 0.12~dex (or around 30~\%).

4) A significant concentration of the most metal-poor dwarf
galaxies with M$_{\rm B}$ ranging from $-12$ to $-15.5$ mag is
discovered in the small cell of the Local Supercluster, which
includes the Lynx-Cancer void. This provides the evidence for the
sizable effect of a very rarefied environment on the evolution of
low-mass galaxies.

\section*{Acknowledgements}

The authors thank S.~Kaisin for providing the H$\alpha$ images of
several program galaxies  prior to the publication. The work of
S.~Pustilnik and A.~Tepliakova was partly supported by the RFBR
grants (project nos. 06-02-16617 and 10-02-92650), as well as the
Federal Target Programm Scientific and Scientific-Pedagogical
Cadre of Innovative Russia (state contract no.\,14.740.11.0901).
A.~Kniazev acknowledges the support from the National Research
Foundation of South Africa. We thank A.~Burenkov and A.~Valeev for
the help with observations. We also thank N.~G.~Guseva for the
constructive comments and suggestions that improved the quality of
the paper. The authors acknowledge the spectral and photometric
data used for this study and the related information available in
the SDSS database. The Sloan Digital Sky Survey (SDSS) is a joint
project of the University of Chicago (Fermilab), the Institute for
Advanced Study, the Japan Participation Group, the John Hopkins
University, the Max-Planck-Institute for Astronomy (MPIA), the
Max-Planck-Institute for Astrophysics (MPA), New Mexico State
University, Princeton University, the United States Naval
Observatory, and the University of Washington. The Apache Point
Observatory, site of the SDSS telescopes, is operated by the
Astrophysical Research Consortium (ARC). This research has made
use of the NASA/IPAC Extragalactic Database (NED), which is
operated by the Jet Propulsion Laboratory, California Institute of
Technology, under contract with the National Aeronautics and Space
Administration, and the HyperLeda database, which is operated by
Lyon University.

\clearpage


\begin{figure*}
 \centering
 \includegraphics[angle=-0,width=15cm]{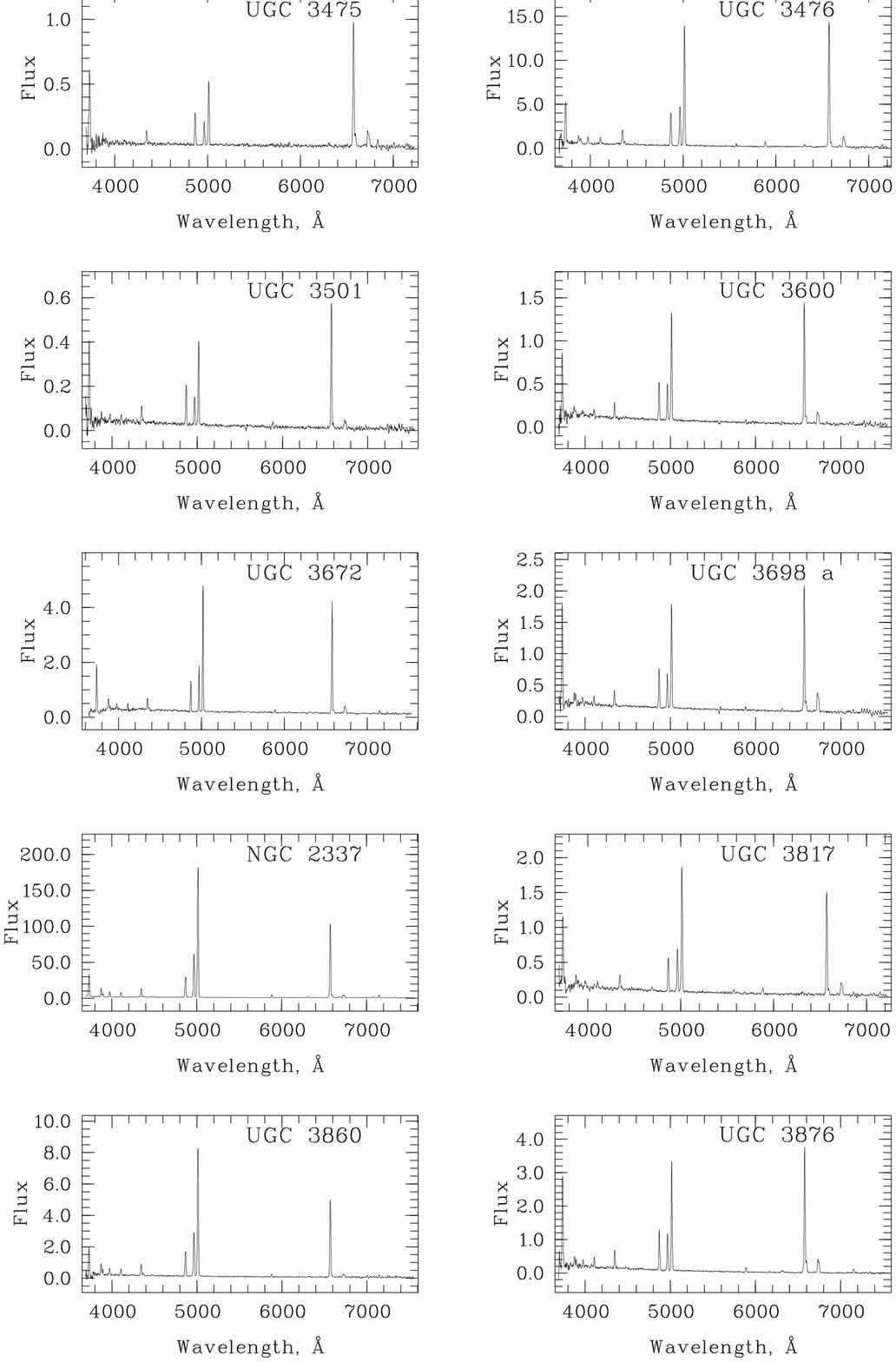}
  \caption{\label{fig:BTA_spectra1} Figure A.1.
Spectra of 10  HII regions in the Lynx-Cancer void galaxies obtained with the
SAO 6m telescope.
}
\end{figure*}

\begin{figure*}
 \centering
 \includegraphics[angle=-0,width=15cm]{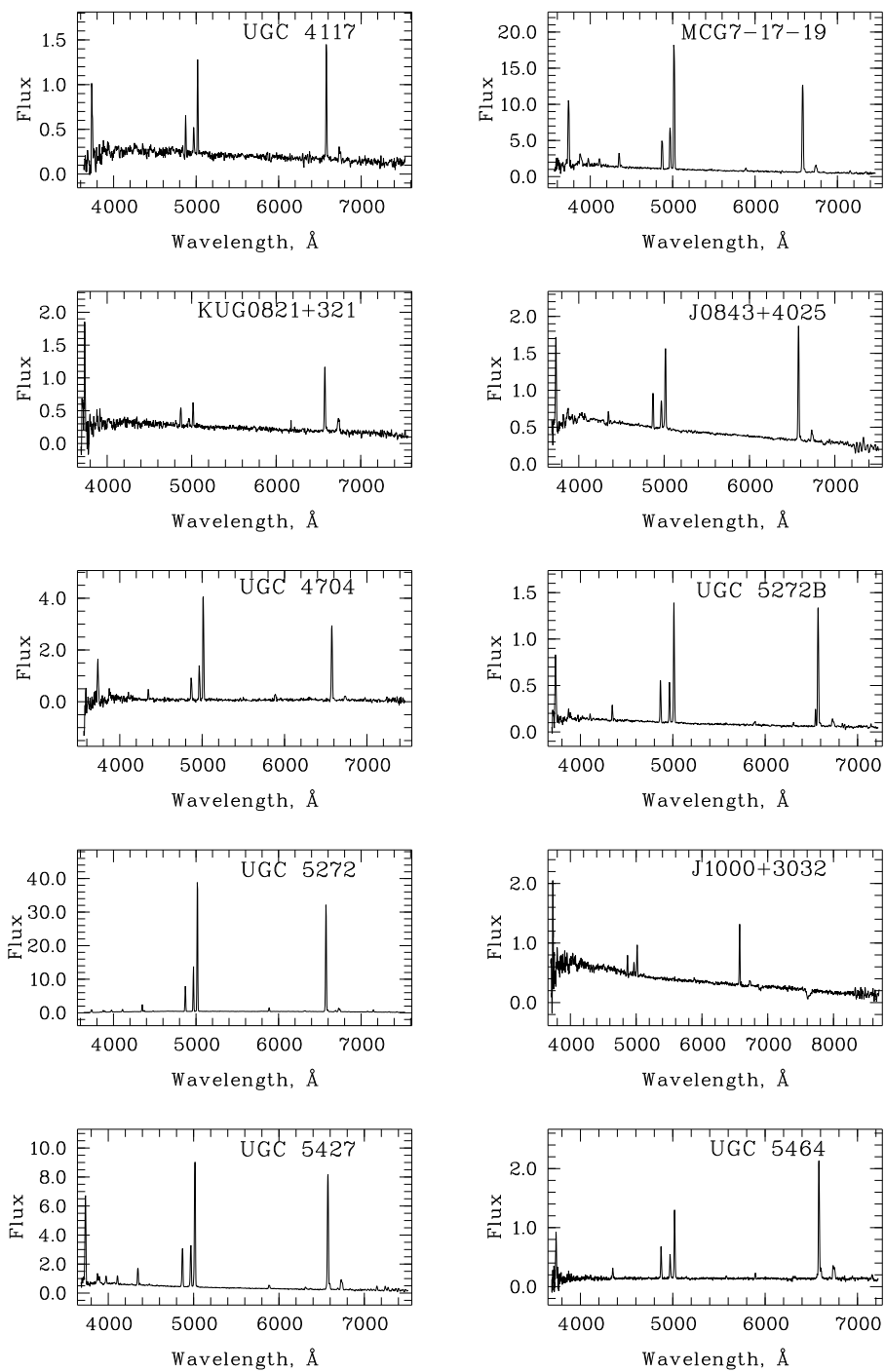}
  \caption{\label{fig:BTA_spectra2} Figure A.2.
Spectra of 10 HII regions in the remaining Lynx-Cancer void galaxies obtained
 with the SAO 6m telescope.
}
\end{figure*}

\begin{figure*}
 \centering
 \includegraphics[angle=-0,width=15cm]{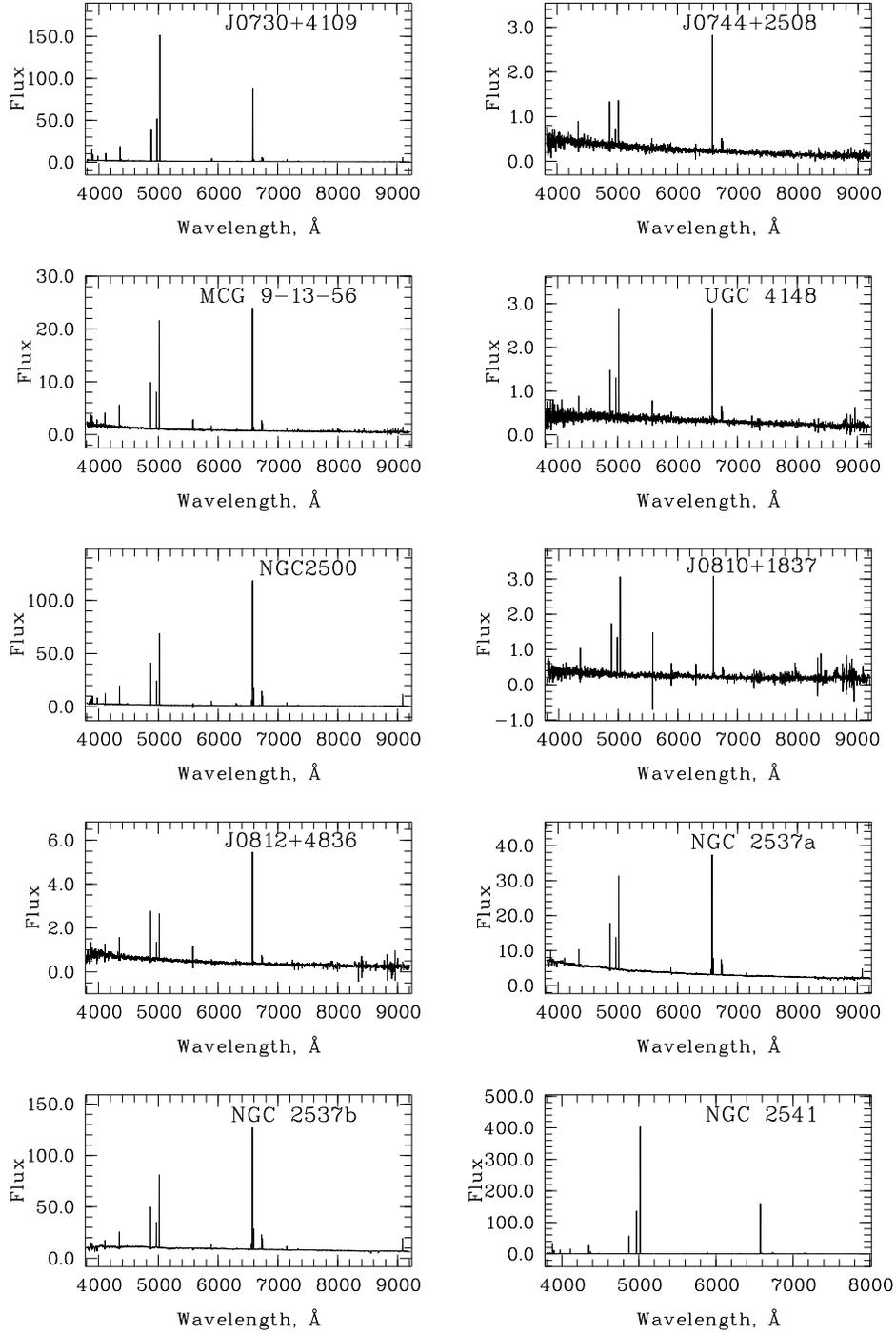}
  \caption{\label{fig:SDSS_spectra3} Figure A.3.
Spectra of 10 HII regions in the Lynx-Cancer void galaxies obtained from
SDSS DR7.
}
\end{figure*}

\begin{figure*}
 \centering
 \includegraphics[angle=-0,width=15cm]{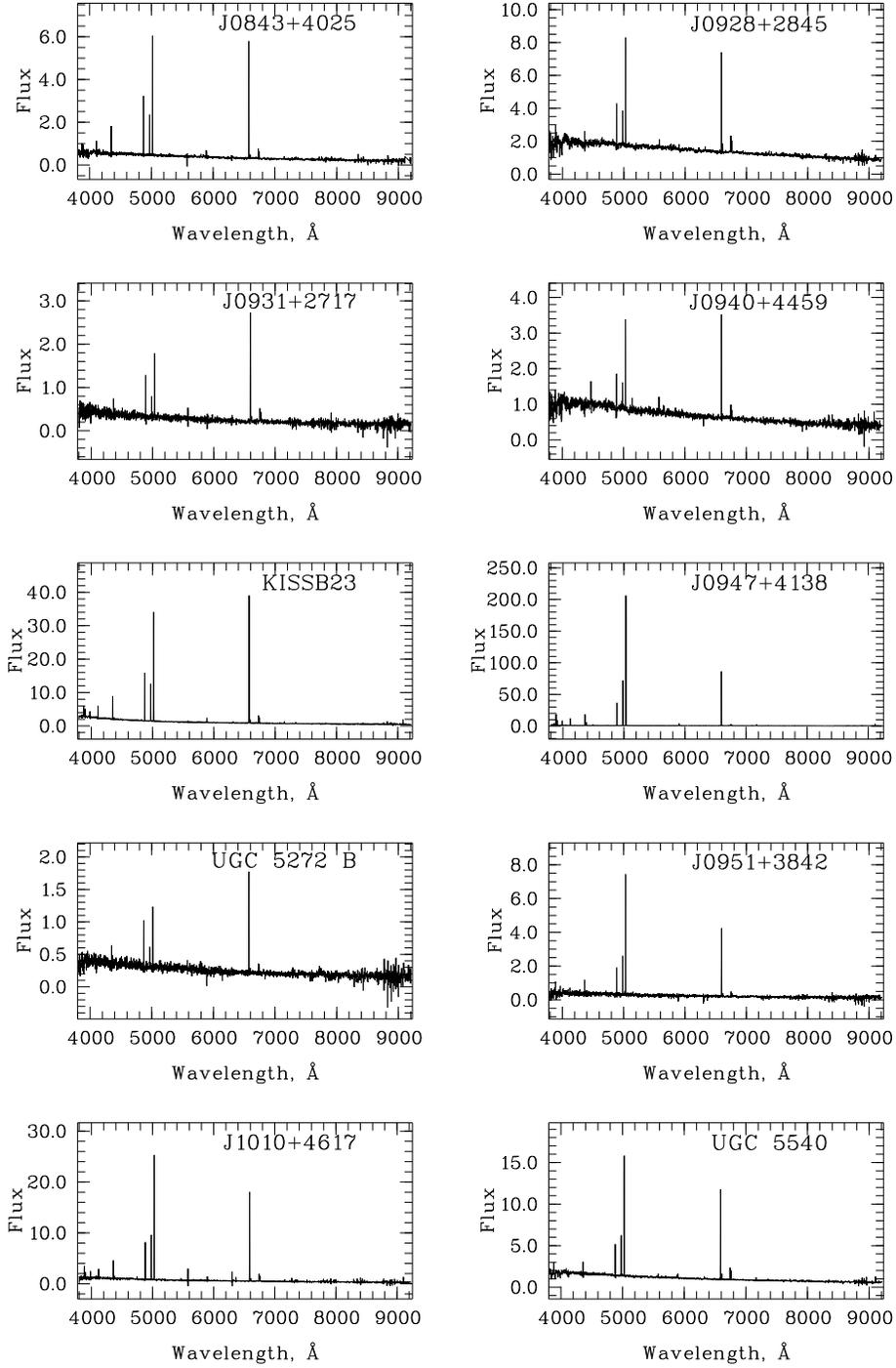}
  \caption{\label{fig:SDSS_spectra4} Figure A.4.
Spectra of other 8 HII regions in the Lynx-Cancer void galaxies obtained
from the SDSS.
}
\end{figure*}

\begin{figure*}
 \centering
 \includegraphics[angle=-0,width=15cm]{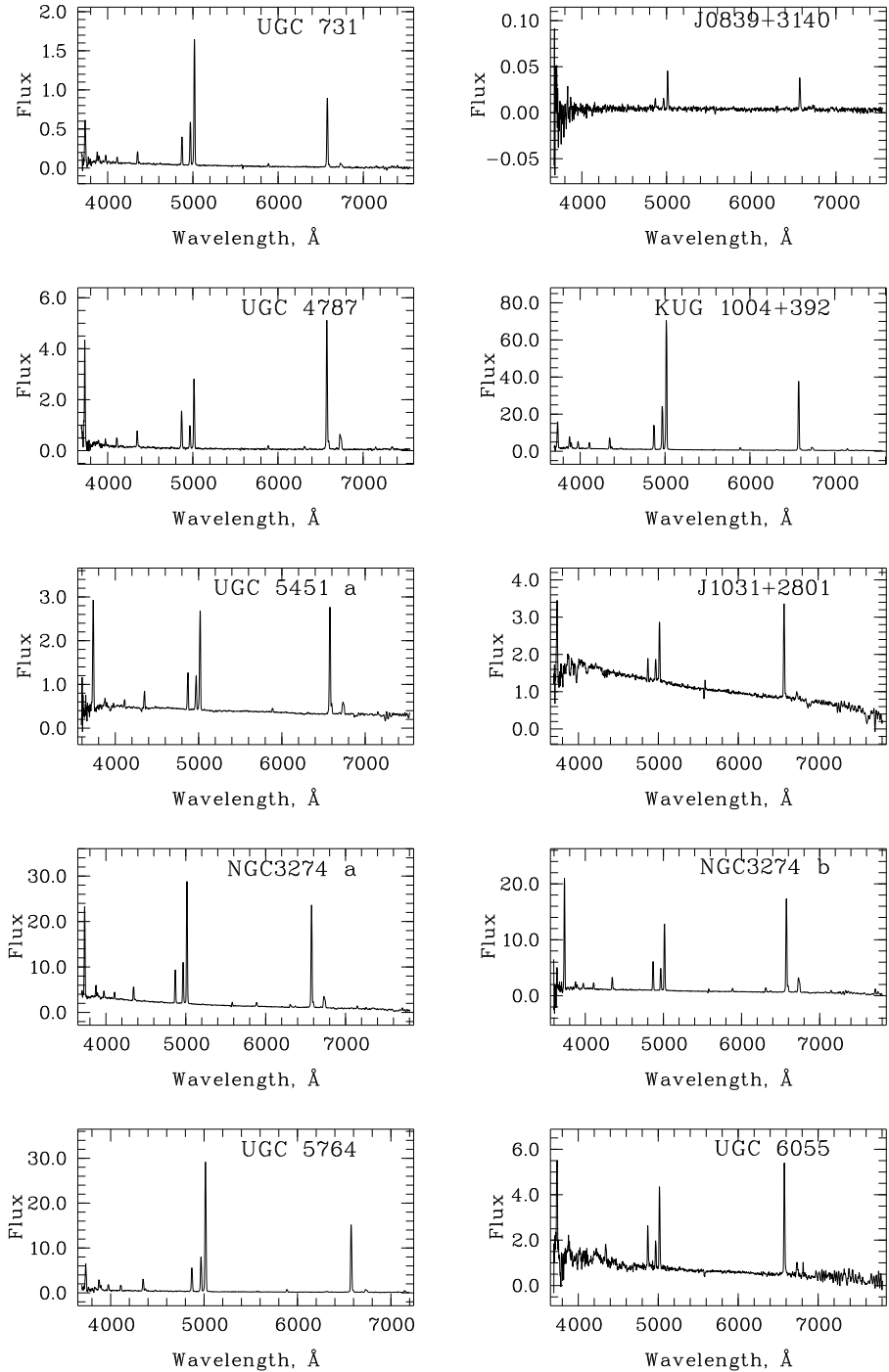}
  \caption{\label{fig:nonvoid_spectra5} Figure A.5.
Spectra of 10  HII regions in 9  galaxies outside the Lynx-Cancer void
obtained with the SAO 6m telescope.
}
\end{figure*}

\clearpage



\renewcommand{\baselinestretch}{1.1}
\begin{table*}
\footnotesize
\begin{flushright}
 {\it APPENDIX B} \\[-5pt]
 \end{flushright} \normalsize
\caption{{\bf Table B.1} Observed and
corrected relative fluxes in the lines of void galaxies (BTA)}
\label{t:Intens_BTA1} 
\begin{tabular}{l|c|c|c|c|c|c} \hline
\rule{0pt}{10pt} & \MC{2}{c|}{UGC3475} & \MC{2}{c|}{UGC3476} &
\MC{2}{c}{UGC3501}     \\ \hline \rule{0pt}{10pt}
$\lambda_{0}$(\AA) Ion                  &
F($\lambda$)/F(H$\beta$)&I($\lambda$)/I(H$\beta$) &
F($\lambda$)/F(H$\beta$)&I($\lambda$)/I(H$\beta$) &
F($\lambda$)/F(H$\beta$)&I($\lambda$)/I(H$\beta$) \\ \hline
3727\ [O\ {\sc ii}]\                           &  2.852$\pm$ 0.235 &  4.460$\pm$ 0.378             &  1.160$\pm$ 0.053 &  1.576$\pm$ 0.074             &  2.216$\pm$ 0.118 &  2.312$\pm$ 0.140            \\
3868\ [Ne\ {\sc iii}]\                         &        ...        &        ...                    &  0.162$\pm$ 0.016 &  0.210$\pm$ 0.021             &  0.168$\pm$ 0.026 &  0.172$\pm$ 0.030            \\
3967\ [Ne\ {\sc iii}]\ +\ H7\                  &        ...        &        ...                    &  0.214$\pm$ 0.013 &  0.275$\pm$ 0.022             &  0.112$\pm$ 0.022 &  0.234$\pm$ 0.055            \\
4101\ H$\delta$\                               &        ...        &        ...                    &  0.198$\pm$ 0.012 &  0.246$\pm$ 0.020             &  0.112$\pm$ 0.018 &  0.228$\pm$ 0.044            \\
4340\ H$\gamma$\                               &  0.426$\pm$ 0.159 &  0.518$\pm$ 0.239             &  0.416$\pm$ 0.017 &  0.478$\pm$ 0.023             &  0.418$\pm$ 0.033 &  0.498$\pm$ 0.044            \\
4363\ [O\ {\sc iii}]\                          &  0.029$\pm$ 0.100 &  0.035$\pm$ 0.121             &  0.068$\pm$ 0.011 &  0.077$\pm$ 0.013             &  0.052$\pm$ 0.022 &  0.050$\pm$ 0.023            \\
\end{tabular}
\end{table*}

\begin{table*}
\caption{{\bf Table B.2}  Observed and
corrected relative fluxes in the lines of void galaxies (BTA)}
\begin{tabular}{l|c|c|c|c|c|c} \hline
& \MC{2}{c|}{UGC3600} & \MC{2}{c|}{UGC3672} & \MC{2}{c}{UGC3698a}
\\ \hline \rule{0pt}{10pt} $\lambda_{0}$(\AA) Ion
& F($\lambda$)/F(H$\beta$)&I($\lambda$)/I(H$\beta$) &
F($\lambda$)/F(H$\beta$)&I($\lambda$)/I(H$\beta$) &
F($\lambda$)/F(H$\beta$)&I($\lambda$)/I(H$\beta$) \\ \hline
3727\ [O\ {\sc ii}]\                           &  1.496$\pm$ 0.099 &  1.636$\pm$ 0.116             &  1.893$\pm$ 0.089 &  2.848$\pm$ 0.145             &  2.836$\pm$ 0.105 &  3.177$\pm$ 0.132            \\
3868\ [Ne\ {\sc iii}]\                         &  0.261$\pm$ 0.027 &  0.280$\pm$ 0.031             &  0.442$\pm$ 0.041 &  0.627$\pm$ 0.060             &  0.254$\pm$ 0.025 &  0.278$\pm$ 0.029            \\
3967\ [Ne\ {\sc iii}]\ +\ H7\                  &  0.140$\pm$ 0.028 &  0.235$\pm$ 0.056             &  0.186$\pm$ 0.026 &  0.259$\pm$ 0.055             &  0.106$\pm$ 0.011 &  0.208$\pm$ 0.026            \\
4101\ H$\delta$\                               &  0.130$\pm$ 0.022 &  0.218$\pm$ 0.044             &  0.222$\pm$ 0.028 &  0.292$\pm$ 0.049             &  0.169$\pm$ 0.009 &  0.261$\pm$ 0.018            \\
4340\ H$\gamma$\                               &  0.445$\pm$ 0.027 &  0.507$\pm$ 0.035             &  0.408$\pm$ 0.030 &  0.488$\pm$ 0.049             &  0.400$\pm$ 0.015 &  0.473$\pm$ 0.020            \\
4363\ [O\ {\sc iii}]\                          &  0.051$\pm$ 0.018 &  0.051$\pm$ 0.019             &  0.064$\pm$ 0.018 &  0.076$\pm$ 0.022             &  0.057$\pm$ 0.008 &  0.058$\pm$ 0.009            \\
\end{tabular}
\end{table*}

\begin{table*}
\caption{{\bf Table B.3} Observed and
corrected relative fluxes in the lines of void galaxies (BTA)}
\label{t:Intens_BTA2}
\begin{tabular}{l|c|c|c|c|c|c} \hline
\rule{0pt}{10pt} & \MC{2}{c|}{UGC3698b} & \MC{2}{c|}{NGC2337} &
\MC{2}{c}{UGC3817}     \\ \hline \rule{0pt}{10pt}
$\lambda_{0}$(\AA) Ion                  &
F($\lambda$)/F(H$\beta$)&I($\lambda$)/I(H$\beta$) &
F($\lambda$)/F(H$\beta$)&I($\lambda$)/I(H$\beta$) &
F($\lambda$)/F(H$\beta$)&I($\lambda$)/I(H$\beta$)  \\ \hline
3727\ [O\ {\sc ii}]\                           &  3.361$\pm$ 0.240 &  4.082$\pm$ 0.341             &  1.054$\pm$ 0.029 &  1.270$\pm$ 0.038             &  2.174$\pm$ 0.125 &  2.406$\pm$ 0.152           \\
3868\ [Ne\ {\sc iii}]\                         &  0.426$\pm$ 0.063 &  0.496$\pm$ 0.083             &  0.414$\pm$ 0.011 &  0.485$\pm$ 0.014             &  0.335$\pm$ 0.084 &  0.362$\pm$ 0.096           \\
3967\ [Ne\ {\sc iii}]\ +\ H7\                  &        ...        &        ...                    &  0.251$\pm$ 0.007 &  0.318$\pm$ 0.012             &  0.171$\pm$ 0.070 &  0.274$\pm$ 0.130           \\
4101\ H$\delta$\                               &  0.118$\pm$ 0.030 &  0.274$\pm$ 0.086             &  0.203$\pm$ 0.007 &  0.257$\pm$ 0.011             &  0.081$\pm$ 0.049 &  0.175$\pm$ 0.129           \\
4340\ H$\gamma$\                               &  0.355$\pm$ 0.045 &  0.479$\pm$ 0.070             &  0.413$\pm$ 0.013 &  0.467$\pm$ 0.016             &  0.498$\pm$ 0.032 &  0.568$\pm$ 0.044           \\
4363\ [O\ {\sc iii}]\                          &  0.106$\pm$ 0.036 &  0.109$\pm$ 0.041             &  0.043$\pm$ 0.006 &  0.045$\pm$ 0.007             &  0.063$\pm$ 0.020 &  0.064$\pm$ 0.021           \\
\end{tabular}
\end{table*}

\begin{table*}
\caption{{\bf Table B.4} Observed and
corrected relative fluxes in the lines of void galaxies (BTA)}
\begin{tabular}{l|c|c|c|c|c|c}
 \hline
& \MC{2}{c|}{UGC3860} & \MC{2}{c|}{UGC3876} & \MC{2}{c}{UGC4117}
\\ \hline \rule{0pt}{10pt} $\lambda_{0}$(\AA) Ion
& F($\lambda$)/F(H$\beta$)&I($\lambda$)/I(H$\beta$) &
F($\lambda$)/F(H$\beta$)&I($\lambda$)/I(H$\beta$) &
F($\lambda$)/F(H$\beta$)&I($\lambda$)/I(H$\beta$) \\ \hline
3727\ [O\ {\sc ii}]\                           &  1.029$\pm$ 0.059 &  1.147$\pm$ 0.067             &  2.235$\pm$ 0.098 &  2.514$\pm$ 0.115             &  2.902$\pm$ 0.300 &  3.466$\pm$ 0.384            \\
3868\ [Ne\ {\sc iii}]\                         &  0.451$\pm$ 0.019 &  0.495$\pm$ 0.021             &  0.199$\pm$ 0.017 &  0.220$\pm$ 0.019             &        ...        &        ...                   \\
3967\ [Ne\ {\sc iii}]\ +\ H7\                  &  0.222$\pm$ 0.016 &  0.243$\pm$ 0.022             &  0.160$\pm$ 0.015 &  0.189$\pm$ 0.022             &        ...        &        ...                   \\
4101\ H$\delta$\                               &  0.222$\pm$ 0.012 &  0.240$\pm$ 0.017             &  0.232$\pm$ 0.013 &  0.262$\pm$ 0.020             &        ...        &        ...                   \\
4340\ H$\gamma$\                               &  0.482$\pm$ 0.021 &  0.506$\pm$ 0.024             &  0.438$\pm$ 0.018 &  0.469$\pm$ 0.022             &        ...        &        ...                   \\
4363\ [O\ {\sc iii}]\                          &  0.106$\pm$ 0.014 &  0.110$\pm$ 0.015             &  0.034$\pm$ 0.012 &  0.035$\pm$ 0.013             &        ...        &        ...                   \\
\end{tabular}
\end{table*}

\begin{table*}
\caption{{\bf Table B.5} Observed and
corrected relative fluxes in the lines of void galaxies (BTA)}
\label{t:Intens_BTA3}
\begin{tabular}{l|c|c|c|c|c|c} \hline
\rule{0pt}{10pt} & \MC{2}{c|}{MCG7-17-19} &
\MC{2}{c|}{KUG0821+321} & \MC{2}{c}{J0843+4025}     \\ \hline
\rule{0pt}{10pt} $\lambda_{0}$(\AA) Ion                  &
F($\lambda$)/F(H$\beta$)&I($\lambda$)/I(H$\beta$) &
F($\lambda$)/F(H$\beta$)&I($\lambda$)/I(H$\beta$) &
F($\lambda$)/F(H$\beta$)&I($\lambda$)/I(H$\beta$) \\ \hline
3727\ [O\ {\sc ii}]\                           &  2.401$\pm$ 0.105 &  2.840$\pm$ 0.133             &  3.548$\pm$ 0.467 &  3.509$\pm$ 0.564             &  2.244$\pm$ 0.092 &  1.616$\pm$ 0.099            \\
3868\ [Ne\ {\sc iii}]\                         &  0.638$\pm$ 0.060 &  0.736$\pm$ 0.070             &        ...        &        ...                    &        ...        &        ...                   \\
3967\ [Ne\ {\sc iii}]\ +\ H7\                  &  0.274$\pm$ 0.036 &  0.316$\pm$ 0.053             &        ...        &        ...                    &        ...        &        ...                   \\
4101\ H$\delta$\                               &  0.258$\pm$ 0.031 &  0.290$\pm$ 0.048             &        ...        &        ...                    &        ...        &        ...                   \\
4340\ H$\gamma$\                               &  0.451$\pm$ 0.024 &  0.487$\pm$ 0.038             &        ...        &        ...                    &  0.198$\pm$ 0.020 &  0.470$\pm$ 0.079            \\
4363\ [O\ {\sc iii}]\                          &  0.081$\pm$ 0.021 &  0.086$\pm$ 0.023             &        ...        &        ...                    &  0.047$\pm$ 0.019 &  0.034$\pm$ 0.019            \\
\end{tabular}
\end{table*}

\begin{table*}
\caption{{\bf Table B.6} Observed and
corrected relative fluxes in the lines of void galaxies (BTA)}
\begin{tabular}{l|c|c|c|c|c|c} \hline
\rule{0pt}{10pt} & \MC{2}{c|}{UGC4704} & \MC{2}{c|}{UGC5272B} &
\MC{2}{c}{UGC5272}     \\ \hline \rule{0pt}{10pt}
$\lambda_{0}$(\AA) Ion                  &
F($\lambda$)/F(H$\beta$)&I($\lambda$)/I(H$\beta$) &
F($\lambda$)/F(H$\beta$)&I($\lambda$)/I(H$\beta$) &
F($\lambda$)/F(H$\beta$)&I($\lambda$)/I(H$\beta$) \\ \hline
3727\ [O\ {\sc ii}]\                           &  1.927$\pm$ 0.165 &  2.252$\pm$ 0.206             &  1.585$\pm$ 0.108 &  1.543$\pm$ 0.124             &  1.209$\pm$ 0.026 &  1.503$\pm$ 0.039            \\
3868\ [Ne\ {\sc iii}]\                         &        ...        &        ...                    &  0.253$\pm$ 0.079 &  0.242$\pm$ 0.087             &  0.397$\pm$ 0.022 &  0.479$\pm$ 0.027            \\
3967\ [Ne\ {\sc iii}]\ +\ H7\                  &        ...        &        ...                    &        ...        &        ...                    &  0.268$\pm$ 0.010 &  0.320$\pm$ 0.015            \\
4101\ H$\delta$\                               &        ...        &        ...                    &  0.078$\pm$ 0.008 &  0.251$\pm$ 0.038             &  0.245$\pm$ 0.009 &  0.285$\pm$ 0.014            \\
4340\ H$\gamma$\                               &  0.403$\pm$ 0.039 &  0.459$\pm$ 0.058             &  0.367$\pm$ 0.021 &  0.481$\pm$ 0.033             &  0.471$\pm$ 0.014 &  0.519$\pm$ 0.017            \\
4363\ [O\ {\sc iii}]\                          &  0.058$\pm$ 0.026 &  0.061$\pm$ 0.028             &  0.054$\pm$ 0.014 &  0.049$\pm$ 0.015             &  0.081$\pm$ 0.009 &  0.088$\pm$ 0.010            \\
\end{tabular}
\end{table*}

\begin{table*}
\caption{{\bf Table B.7} Observed and
corrected relative fluxes in the lines of void galaxies   (BTA) }
\label{t:Intens_BTA4}
\begin{tabular}{l|c|c|c|c|c|c} \hline
\rule{0pt}{10pt} & \MC{2}{c|}{J1000+3032} & \MC{2|}{c}{UGC5427} &
\MC{2}{c}{UGC5464}     \\ \hline \rule{0pt}{10pt}
$\lambda_{0}$(\AA) Ion                  &
F($\lambda$)/F(H$\beta$)&I($\lambda$)/I(H$\beta$) &
F($\lambda$)/F(H$\beta$)&I($\lambda$)/I(H$\beta$) &
F($\lambda$)/F(H$\beta$)&I($\lambda$)/I(H$\beta$) \\ \hline
3727\ [O\ {\sc ii}]\                           &  2.574$\pm$ 0.393 &  1.963$\pm$ 0.431             &  2.193$\pm$ 0.066 &  2.202$\pm$ 0.073             &  1.451$\pm$ 0.078 &  2.224$\pm$ 0.123            \\
3967\ [Ne\ {\sc iii}]\ +\ H7\                  &        ...        &        ...                    &  0.180$\pm$ 0.008 &  0.251$\pm$ 0.015             &        ...        &        ...                   \\
4101\ H$\delta$\                               &        ...        &        ...                    &  0.187$\pm$ 0.008 &  0.252$\pm$ 0.014             &        ...        &        ...                   \\
4340\ H$\gamma$\                               &        ...        &        ...                    &  0.422$\pm$ 0.014 &  0.471$\pm$ 0.018             &  0.388$\pm$ 0.022 &  0.470$\pm$ 0.034            \\
4363\ [O\ {\sc iii}]\                          &        ...        &        ...                    &  0.040$\pm$ 0.008 &  0.039$\pm$ 0.008             &  0.023$\pm$ 0.012 &  0.027$\pm$ 0.015            \\
4861\ H$\beta$\                                &  1.000$\pm$ 0.123 &  1.000$\pm$ 0.190             &  1.000$\pm$ 0.030 &  1.000$\pm$ 0.032             &  1.000$\pm$ 0.030 &  1.000$\pm$ 0.036            \\
\end{tabular}
\end{table*}

\begin{table*}
\caption{{\bf Table B.8} Observed and
corrected relative fluxes in the lines of void galaxies (SDSS) }
\label{t:Intens_SDSS1}
\begin{tabular}{l|c|c|c|c|c|c} \hline
\rule{0pt}{10pt} & \MC{2}{c|}{J0730+4109} & \MC{2}{c|}{J0744+2508}
& \MC{2}{c}{MCG9-13-56}     \\ \hline \rule{0pt}{10pt}
$\lambda_{0}$(\AA) Ion                  &
F($\lambda$)/F(H$\beta$)&I($\lambda$)/I(H$\beta$) &
F($\lambda$)/F(H$\beta$)&I($\lambda$)/I(H$\beta$) &
F($\lambda$)/F(H$\beta$)&I($\lambda$)/I(H$\beta$)  \\ \hline
(3727\ [O\ {\sc ii}])                           &        ...        &        ...                    &  1.143$\pm$ 0.865 &  0.946$\pm$ 0.963             &  1.905$\pm$ 0.235 &  2.043$\pm$ 0.264            \\
3868\ [Ne\ {\sc iii}]\                         &  0.347$\pm$ 0.014 &  0.408$\pm$ 0.017             &        ...        &        ...                    &  0.185$\pm$ 0.045 &  0.196$\pm$ 0.050            \\
4101\ H$\delta$\                               &  0.220$\pm$ 0.010 &  0.253$\pm$ 0.015             &        ...        &        ...                    &  0.210$\pm$ 0.017 &  0.280$\pm$ 0.030            \\
4340\ H$\gamma$\                               &  0.457$\pm$ 0.014 &  0.498$\pm$ 0.018             &  0.210$\pm$ 0.061 &  0.470$\pm$ 0.220             &  0.407$\pm$ 0.022 &  0.466$\pm$ 0.031            \\
4363\ [O\ {\sc iii}]\                          &  0.054$\pm$ 0.007 &  0.058$\pm$ 0.008             &        ...        &        ...                    &        ...        &        ...                   \\
\end{tabular}
\end{table*}

\begin{table*}
\caption{{\bf Table B.9} Observed and
corrected relative fluxes in the lines of void galaxies (SDSS) }
\label{t:Intens_SDSS2}
\begin{tabular}{l|c|c|c|c|c|c} \hline
\rule{0pt}{10pt} & \MC{2}{c|}{UGC4148} & \MC{2}{c|}{NGC2500} &
\MC{2}{c}{MCG7-17-19}     \\ \hline \rule{0pt}{10pt}
$\lambda_{0}$(\AA) Ion                  &
F($\lambda$)/F(H$\beta$)&I($\lambda$)/I(H$\beta$) &
F($\lambda$)/F(H$\beta$)&I($\lambda$)/I(H$\beta$) &
F($\lambda$)/F(H$\beta$)&I($\lambda$)/I(H$\beta$) \\ \hline
(3727\ [O\ {\sc ii}])                          &        ...        &        ...                    &  0.861$\pm$ 0.087 &  1.140$\pm$ 0.117             &        ...        &        ...                  \\
3868\ [Ne\ {\sc iii}]\                         &  ...              &  0.235$\pm$ 0.209             &  0.145$\pm$ 0.024 &  0.184$\pm$ 0.030             &  0.257$\pm$ 0.014 &  0.274$\pm$ 0.015           \\
4101\ H$\delta$\                               &        ...        &        ...                    &  0.218$\pm$ 0.009 &  0.267$\pm$ 0.014             &  0.222$\pm$ 0.011 &  0.257$\pm$ 0.015           \\
4340\ H$\gamma$\                               &  0.340$\pm$ 0.119 &  0.477$\pm$ 0.207             &  0.430$\pm$ 0.011 &  0.489$\pm$ 0.014             &  0.439$\pm$ 0.013 &  0.471$\pm$ 0.016           \\
4363\ [O\ {\sc iii}]\                          &        ...        &        ...                    &        ...        &        ...                    &  0.044$\pm$ 0.007 &  0.045$\pm$ 0.008           \\
\end{tabular}
\end{table*}

\begin{table*}
 \caption{{\bf Table B.10} Observed and
corrected relative fluxes in the lines of void galaxies (SDSS) }
\begin{tabular}{l|c|c|c|c|c|c} \hline
\rule{0pt}{10pt} & \MC{2}{c|}{J0810+1837} & \MC{2}{c|}{J0812+4836}
& \MC{2}{c}{NGC2537}     \\ \hline \rule{0pt}{10pt}
$\lambda_{0}$(\AA) Ion                  &
F($\lambda$)/F(H$\beta$)&I($\lambda$)/I(H$\beta$) &
F($\lambda$)/F(H$\beta$)&I($\lambda$)/I(H$\beta$) &
F($\lambda$)/F(H$\beta$)&I($\lambda$)/I(H$\beta$)  \\ \hline
(3727\ [O\ {\sc ii}])                           &  4.040$\pm$ 2.070 &  3.652$\pm$ 2.174             &  ...              &  ...                          &  ...              &  ...                         \\
3868\ [Ne\ {\sc iii}]\                         &        ...        &        ...                    &  0.417$\pm$ 0.155 &  0.346$\pm$ 0.155             &  0.129$\pm$ 0.045 &  0.126$\pm$ 0.053            \\
4101\ H$\delta$\                               &  0.075$\pm$ 0.067 &  0.269$\pm$ 0.343             &  0.080$\pm$ 0.044 &  0.294$\pm$ 0.260             &  0.070$\pm$ 0.024 &  0.285$\pm$ 0.154            \\
4340\ H$\gamma$\                               &  0.376$\pm$ 0.073 &  0.478$\pm$ 0.122             &  0.266$\pm$ 0.046 &  0.427$\pm$ 0.118             &  0.286$\pm$ 0.026 &  0.462$\pm$ 0.062            \\
4363\ [O\ {\sc iii}]\                          &        ...        &        ...                    &        ...        &        ...                    &  0.066$\pm$ 0.020 &  0.059$\pm$ 0.022            \\
\end{tabular}
\end{table*}

\begin{table*}
\caption{{\bf Table B.11} Observed and
corrected relative fluxes in the lines of void galaxies (SDSS) }
\label{t:Intens_SDSS3}
\begin{tabular}{l|c|c|c|c|c|c} \hline
\rule{0pt}{10pt} & \MC{2}{c|}{NGC2541} & \MC{2}{c|}{J0843+4025} &
\MC{2}{c}{J0911+3135}     \\ \hline \rule{0pt}{10pt}
$\lambda_{0}$(\AA) Ion                  &
F($\lambda$)/F(H$\beta$)&I($\lambda$)/I(H$\beta$) &
F($\lambda$)/F(H$\beta$)&I($\lambda$)/I(H$\beta$) &
F($\lambda$)/F(H$\beta$)&I($\lambda$)/I(H$\beta$) \\ \hline
(3727\ [O\ {\sc ii}])                           &        ...        &        ...                    &  2.110$\pm$ 0.534 &  1.947$\pm$ 0.552             &  2.099$\pm$ 0.892 &  2.081$\pm$ 1.025            \\
3868\ [Ne\ {\sc iii}]\                         &  0.516$\pm$ 0.015 &  0.656$\pm$ 0.020             &  0.257$\pm$ 0.093 &  0.236$\pm$ 0.095             &        ...        &        ...                   \\
4101\ H$\delta$\                               &  0.231$\pm$ 0.007 &  0.282$\pm$ 0.011             &  0.119$\pm$ 0.031 &  0.243$\pm$ 0.086             &        ...        &        ...                   \\
4340\ H$\gamma$\                               &  0.416$\pm$ 0.011 &  0.473$\pm$ 0.013             &  0.406$\pm$ 0.029 &  0.487$\pm$ 0.049             &  0.332$\pm$ 0.039 &  0.476$\pm$ 0.077            \\
4740\ [Ar\ {\sc iv]}\                          &  0.009$\pm$ 0.002 &  0.009$\pm$ 0.002             &        ...        &        ...                    &        ...        &        ...                   \\
\end{tabular}
\end{table*}

\begin{table*}
 \caption{{\bf Table B.12} Observed and
corrected relative fluxes in the lines of void galaxies (SDSS) }
\begin{tabular}{l|c|c|c|c|c|c} \hline
\rule{0pt}{10pt} & \MC{2}{c|}{J0928+2845} & \MC{2}{c|}{J0931+2717}
& \MC{2}{c}{J0940+4459}     \\ \hline \rule{0pt}{10pt}
$\lambda_{0}$(\AA) Ion                  &
F($\lambda$)/F(H$\beta$)&I($\lambda$)/I(H$\beta$) &
F($\lambda$)/F(H$\beta$)&I($\lambda$)/I(H$\beta$) &
F($\lambda$)/F(H$\beta$)&I($\lambda$)/I(H$\beta$)  \\ \hline
(3727\ [O\ {\sc ii}])                           &        ...        &        ...                    &  1.521$\pm$ 0.364 &  1.657$\pm$ 0.490   &  1.503$\pm$ 0.353 &  1.327$\pm$ 0.370                     \\
3868\ [Ne\ {\sc iii}]\                         &  1.009$\pm$ 0.327 &  0.634$\pm$ 0.332             &        ...        &        ...           &        ...        &        ...                            \\
4340\ H$\gamma$\                               &  0.108$\pm$ 0.070 &  0.450$\pm$ 0.613             &  0.270$\pm$ 0.073 &  0.469$\pm$ 0.184    &  0.259$\pm$ 0.092 &  0.357$\pm$ 0.169                     \\
4861\ H$\beta$\                                &  1.000$\pm$ 0.172 &  1.000$\pm$ 0.292             &  1.000$\pm$ 0.133 &  1.000$\pm$ 0.173    &  1.000$\pm$ 0.252 &  1.000$\pm$ 0.295                     \\
4959\ [O\ {\sc iii}]\                          &  1.206$\pm$ 0.165 &  0.753$\pm$ 0.165             &  0.494$\pm$ 0.082 &  0.400$\pm$ 0.080    &  0.674$\pm$ 0.181 &  0.595$\pm$ 0.181                     \\
5007\ [O\ {\sc iii}]\                          &  3.639$\pm$ 0.454 &  2.271$\pm$ 0.454             &  2.128$\pm$ 0.216 &  1.706$\pm$ 0.211    &  2.044$\pm$ 0.392 &  1.804$\pm$ 0.393                     \\
\end{tabular}
\end{table*}

\begin{table*}
\caption{{\bf Table B.13} Observed and
corrected relative fluxes in the lines of void galaxies (SDSS) }
\label{t:Intens_SDSS4}
\begin{tabular}{l|c|c|c|c|c|c} \hline
\rule{0pt}{10pt} & \MC{2}{c|}{KISSB23} & \MC{2}{c|}{J0947+4138} &
\MC{2}{c}{J0951+3842}     \\ \hline \rule{0pt}{10pt}
$\lambda_{0}$(\AA) Ion                  &
F($\lambda$)/F(H$\beta$)&I($\lambda$)/I(H$\beta$) &
F($\lambda$)/F(H$\beta$)&I($\lambda$)/I(H$\beta$) &
F($\lambda$)/F(H$\beta$)&I($\lambda$)/I(H$\beta$) \\ \hline
(3727\ [O\ {\sc ii}])                           &  1.682$\pm$ 0.215 &  1.651$\pm$ 0.225             &        ...        &        ...                    &        ...        &        ...                   \\
3868\ [Ne\ {\sc iii}]\                         &  0.191$\pm$ 0.020 &  0.187$\pm$ 0.021             &  0.452$\pm$ 0.014 &  0.497$\pm$ 0.016             &        ...        &        ...                   \\
4101\ H$\delta$\                               &  0.186$\pm$ 0.012 &  0.275$\pm$ 0.024             &  0.258$\pm$ 0.008 &  0.279$\pm$ 0.012             &  0.070$\pm$ 0.059 &  0.247$\pm$ 0.311            \\
4340\ H$\gamma$\                               &  0.409$\pm$ 0.013 &  0.473$\pm$ 0.019             &  0.476$\pm$ 0.011 &  0.501$\pm$ 0.014             &  0.304$\pm$ 0.081 &  0.480$\pm$ 0.176            \\
4363\ [O\ {\sc iii}]\                          &        ...        &        ...                    &  0.122$\pm$ 0.007 &  0.127$\pm$ 0.008             &        ...        &        ...                   \\
\end{tabular}
\end{table*}

\begin{table*}
\caption{{\bf Table B.14} Observed and
corrected relative fluxes in the lines of void galaxies  (SDSS) }
\begin{tabular}{l|c|c|c|c}
 \hline \rule{0pt}{10pt} & \MC{2}{c|}{J1000+3032} &
\MC{2}{c}{J1010+4617}     \\
\hline \rule{0pt}{10pt} $\lambda_{0}$(\AA) Ion                  &
F($\lambda$)/F(H$\beta$)&I($\lambda$)/I(H$\beta$) &
F($\lambda$)/F(H$\beta$)&I($\lambda$)/I(H$\beta$)  \\ \hline

3868\ [Ne\ {\sc iii}]\                         &        ...        &        ...                    &  0.344$\pm$ 0.054 &  0.345$\pm$ 0.059            \\
4101\ H$\delta$\                               &        ...        &        ...                    &  0.119$\pm$ 0.026 &  0.216$\pm$ 0.062            \\
4340\ H$\gamma$\                               &  0.265$\pm$ 0.141 &  0.471$\pm$ 0.377             &  0.430$\pm$ 0.029 &  0.499$\pm$ 0.043            \\
4861\ H$\beta$\                                &  1.000$\pm$ 0.166 &  1.000$\pm$ 0.235             &  1.000$\pm$ 0.037 &  1.000$\pm$ 0.044            \\
4959\ [O\ {\sc iii}]\                          &  0.476$\pm$ 0.108 &  0.379$\pm$ 0.108             &  1.075$\pm$ 0.042 &  0.994$\pm$ 0.042            \\
5007\ [O\ {\sc iii}]\                          &  1.386$\pm$ 0.199 &  1.100$\pm$ 0.197             &  3.306$\pm$ 0.118 &  3.048$\pm$ 0.117            \\
\end{tabular}
\end{table*}

\clearpage

\begin{table*}
\caption{{\bf Table B.15} Oxygen
abundances in the studied void galaxies   (BTA)}
\label{t:Chem_BTA}
\begin{tabular}{l|c|c|c|c|c|c} \hline
\rule{0pt}{10pt} Value                               & UGC3475 &
UGC3476a           & UGC3501           & UGC3600           &
UGC3672           & UGC3698a              \\ \hline
$T_{\rm e}$(OIII)(K)\                & 14,533$\pm$1082~~ & 15,773$\pm$1240~~  & 17,155$\pm$1051~~ & 16,404$\pm$1037~~ & 14,812$\pm$1950~~ & 16,319$\pm$1261~~     \\
$T_{\rm e}$(OII)(K)\                 & 13,654$\pm$1243~~ & 14,261$\pm$1341~~  & 14,716$\pm$1082~~ & 14,498$\pm$1094~~ & 13,807$\pm$2208~~ & 14,469$\pm$1335~~     \\
O$^{+}$/H$^{+}$($\times$10$^5$)\     & 5.514$\pm$1.739~~ & 1.675$\pm$0.513~~  & 2.278$\pm$0.554~~ & 1.677$\pm$0.420~~ & 3.364$\pm$1.785~~ & 3.224$\pm$0.956~~     \\
O$^{++}$/H$^{+}$($\times$10$^5$)\    & 2.414$\pm$0.471~~ & 3.462$\pm$0.665~~  & 1.385$\pm$0.201~~ & 2.387$\pm$0.362~~ & 4.569$\pm$1.540~~ & 2.240$\pm$0.414~~     \\
O/H($\times$10$^5$)\                 & 7.928$\pm$1.802~~ & 5.137$\pm$0.840~~  & 3.664$\pm$0.590~~ & 4.064$\pm$0.554~~ & 7.933$\pm$2.358~~ & 5.464$\pm$1.041~~     \\
12+log(O/H)\                         & ~7.90$\pm$0.10~~  & ~7.71$\pm$0.07~~   & ~7.56$\pm$0.07~~  & ~7.61$\pm$0.06~~  & ~7.90$\pm$0.13~~  & ~7.74$\pm$0.08~~      \\
& & & & & &\\
12+log(O/H)(PVT10)\                   & ~8.23$\pm$0.14~~  & ~7.76$\pm$0.13~~   & ~7.69$\pm$0.17~~  & ~7.86$\pm$0.13~~  & ~7.94$\pm$0.15~~  & ~8.05$\pm$0.11~~      \\
12+log(O/H)(PM10)\                   & ~8.23$\pm$0.09~~  & ~7.74$\pm$0.12~~   & ~7.55$\pm$0.13~~  & ~7.90$\pm$0.10~~  & ~7.98$\pm$0.12~~  & ~7.95$\pm$0.09~~      \\
\hline
\rule{0pt}{10pt} Value                               & UGC3698b &
NGC2337a           & UGC3817            & UGC3860            &
UGC3876             & UGC4117               \\ \hline
$T_{\rm e}$(OIII)(K)\                & 14,847$\pm$1103~~ & 10,475$\pm$504 ~~  & 14,173$\pm$2021~~  & 15,809$\pm$1018~~  & 15,442$\pm$1025~~   & 15,018$\pm$1175~~     \\
$T_{\rm e}$(OII)(K)\                 & 13,826$\pm$1246~~ & 10,441$\pm$504 ~~  & 13,443$\pm$2366~~  & 14,276$\pm$1100~~  & 14,118$\pm$1125~~   & 13,914$\pm$1317~~     \\
O$^{+}$/H$^{+}$($\times$10$^5$)\     & 4.627$\pm$1.429~~ & 4.700$\pm$0.834~~  & 3.133$\pm$1.885~~  & 1.215$\pm$0.309~~  & 2.862$\pm$0.753~~   & 3.991$\pm$1.316~~     \\
O$^{++}$/H$^{+}$($\times$10$^5$)\    & 2.442$\pm$0.479~~ & 19.190$\pm$3.110~~ & 4.769$\pm$1.793~~  & 4.917$\pm$0.777~~  & 2.773$\pm$0.459~~   & 2.598$\pm$0.549~~     \\
O/H($\times$10$^5$)\                 & 7.069$\pm$1.507~~ & 23.890$\pm$3.220~~ & 9.214$\pm$2.688~~  & 6.133$\pm$0.836~~  & 5.636$\pm$0.882~~   & 6.589$\pm$1.426~~     \\
12+log(O/H)\                         & ~7.85$\pm$0.09~~  & ~8.38$\pm$0.06~~   & ~7.96$\pm$0.13~~   & ~7.79$\pm$0.06~~   & ~7.75$\pm$0.07~~    & ~7.82$\pm$0.09~~      \\
& & & & & &\\
12+log(O/H)(PVT10)\                   & ...            ~  & ~7.93$\pm$0.11~~   & ~8.02$\pm$0.15~~   & ~7.78$\pm$0.16~~   & ~8.25$\pm$0.11~~    & ~7.89$\pm$0.26~~      \\
12+log(O/H)(PM10)\                   & ~7.88$\pm$0.11~~  & ~8.06$\pm$0.11~~   & ~8.01$\pm$0.11~~   & ~7.82$\pm$0.18~~   & ~8.26$\pm$0.08~~    & ~7.69$\pm$0.16~~      \\
\hline

Value                               & MCG7-17-19        &
KUG0821+32 & J0843+4025         & UGC4704            & UGC5272B &
UGC5272              \\ \hline
$T_{\rm e}$(OIII)(K)\                & 15,559$\pm$1969~~ & 16,807$\pm$1442~~  & 17,994$\pm$1060~~  & 14,102$\pm$1079~~  & 14,943$\pm$2058~~  & 14,656$\pm$778 ~~    \\
$T_{\rm e}$(OII)(K)\                 & 14,170$\pm$2151~~ & 14,624$\pm$1500~~  & 14,879$\pm$1071~~  & 13,400$\pm$1268~~  & 13,876$\pm$2315~~  & 13,723$\pm$888 ~~    \\
O$^{+}$/H$^{+}$($\times$10$^5$)\     & 3.082$\pm$1.520~~ & 3.444$\pm$1.244~~  & 1.487$\pm$0.345~~  & 2.951$\pm$0.983~~  & 1.794$\pm$0.995~~  & 1.879$\pm$0.339~~    \\
O$^{++}$/H$^{+}$($\times$10$^5$)\    & 4.130$\pm$1.290~~ & 0.874$\pm$0.217~~  & 1.297$\pm$0.178~~  & 5.288$\pm$1.094~~  & 2.847$\pm$0.995~~  & 5.617$\pm$0.775~~    \\
O/H($\times$10$^5$)\                 & 7.213$\pm$1.994~~ & 4.318$\pm$1.263~~  & 2.784$\pm$0.389~~  & 8.238$\pm$1.471~~  & 4.641$\pm$1.407~~  & 7.497$\pm$0.845~~    \\
12+log(O/H)\                         & ~7.86$\pm$0.12~~  & ~7.64$\pm$0.13~~   & ~7.44$\pm$0.06~~   & ~7.92$\pm$0.08~~   & ~7.67$\pm$0.13~~   & ~7.87$\pm$0.05~~     \\
& & & & & &\\
12+log(O/H)(PVT10)\                   &       ...         & ...                & ~7.61$\pm$0.20~~   &       ...          & ~7.66$\pm$0.23~~   & ~7.73$\pm$0.11~~     \\
12+log(O/H)(PM10)\                   &       ...         & ~7.49$\pm$0.24~~   & ~7.49$\pm$0.18~~   &       ...          & ~7.65$\pm$0.26~~   & ~7.78$\pm$0.12~~     \\
\hline
\end{tabular}
\end{table*}

\begin{table*}
\caption{{\bf Table B.16} Oxygen
abundances in the studied void galaxies   (BTA)} \label{t:Chem}
\begin{tabular}{l|c|c|c} \hline
\rule{0pt}{10pt} Value                             & J1000+3032
& UGCC5427           & UGC5464            \\ \hline
$T_{\rm e}$(OIII)(K)\                & 18,389$\pm$1521~~ & 12,429$\pm$1005~~  & 12,399$\pm$2554~~   \\
$T_{\rm e}$(OII)(K)\                 & 14,926$\pm$1528~~ & 12,196$\pm$1300~~  & 12,171$\pm$3312~~   \\
O$^{+}$/H$^{+}$($\times$10$^5$)\     & 1.807$\pm$0.698~~ & 4.005$\pm$1.568~~  & 4.233$\pm$4.230~~   \\
O$^{++}$/H$^{+}$($\times$10$^5$)\    & 0.919$\pm$0.201~~ & 5.813$\pm$1.374~~  & 4.097$\pm$2.461~~   \\
O/H($\times$10$^5$)\                 & 2.726$\pm$0.726~~ & 9.818$\pm$2.085~~  & 8.330$\pm$4.894~~   \\
12+log(O/H)\                         & ~7.44$\pm$0.12~~  & ~7.99$\pm$0.09~~   & ~7.92$\pm$0.26~~    \\
& & &\\
12+log(O/H)(PVT10)\                   & ...               & ~7.91$\pm$0.11~~   & ~8.23$\pm$0.11~~    \\
12+log(O/H)(PM10)\                   & ~7.58$\pm$0.24~~  & ~7.92$\pm$0.10~~   & ~8.24$\pm$0.08~~    \\
\hline
\end{tabular}
\end{table*}

\begin{table*}
\caption{~~~~~~~~~~~~~~~~~~~~~~~~~~~~~{\bf
Table B.17} Oxygen abundances in the studied galaxies (SDSS)}
\label{t:Chem}
\begin{tabular}{l|c|c} \hline
\rule{0pt}{10pt} Value                            & J1000+3032 &
J1010+4617                     \\ \hline
$T_{\rm e}$(OIII)(K)\                & 19,565$\pm$3124~~   & 16,098$\pm$2701~~       \\
$T_{\rm e}$(OII)(K)\                 & 15,569$\pm$3145~~   & 14,389$\pm$1010~~       \\
O$^{+}$/H$^{+}$($\times$10$^5$)\     & ...                 & 2.821$\pm$1.160~~       \\
O$^{++}$/H$^{+}$($\times$10$^5$)\    & 0.663$\pm$0.243~~   & 2.802$\pm$1.127~~       \\
O/H($\times$10$^5$)\                 & $>$0.663$\pm$0.243~~  & 5.623$\pm$1.617~~       \\
12+log(O/H)\                         & $>$6.82$\pm$0.16~~    & ~7.75$\pm$0.12~~        \\
& & \\
12+log(O/H)(PM10)\                   & ~7.34$\pm$0.19~~    & ~7.81$\pm$0.09~~        \\
\hline
\end{tabular}
\end{table*}

\begin{table*}
\caption{{\bf Table B.18} Oxygen
abundances in the studied void galaxies (SDSS)}
\label{t:Chem_SDSS}
\begin{tabular}{l|c|c|c|c|c|c} \hline
\rule{0pt}{10pt} Value                             & J0730+4109 &
J0744+2508          & MCG9-13-56          & UGC~4148 & NGC~2500 &
MCG~7-17-19                  \\ \hline
$T_{\rm e}$(OIII)(K)\                & 12,712$\pm$664 ~~     & 22,040$\pm$3809~~   & 16,538$\pm$1081~~   & 16,343$\pm$2697~~     & 14,370$\pm$2348~~    & 13,280$\pm$924 ~~     \\
$T_{\rm e}$(OII)(K)\                 & 12,424$\pm$522 ~~     & 16,211$\pm$3900~~   & 14,542$\pm$1136~~   & 14,477$\pm$2852~~     & 13,561$\pm$2720~~    & 12,851$\pm$663 ~~     \\
O$^{+}$/H$^{+}$($\times$10$^5$)\     & 3.254$\pm$0.725~~     & 0.721$\pm$0.686~~   & 2.031$\pm$0.454~~   &  7.675$\pm$11.20~~    & 1.758$\pm$1.688~~    & 3.481$\pm$0.976~~     \\
O$^{++}$/H$^{+}$($\times$10$^5$)\    & 7.718$\pm$1.164~~     & 0.384$\pm$0.148~~   & 2.009$\pm$0.311~~   & 1.890$\pm$0.775~~     & 2.097$\pm$0.892~~    & 4.725$\pm$0.913~~     \\
O$^{+++}$/H$^{+}$($\times$10$^5$)\    & ...                   & ...                 & ...                & ...                   & ...                  & 0.126$\pm$0.055~~     \\
O/H($\times$10$^5$)\                 & 10.970$\pm$1.371~~    & 1.105$\pm$0.701~~   & 4.039$\pm$0.551~~   &  9.565$\pm$11.22~~    & 3.856$\pm$1.909~~    & 8.333$\pm$1.338~~     \\
12+log(O/H)\                         & ~8.04$\pm$0.05~~      & ~7.04$\pm$0.28~~    & ~7.61$\pm$0.06~~    & ~7.98$\pm$0.51~~      & ~7.59$\pm$0.21~~     & ~7.92$\pm$0.07~~      \\
& & & & & &\\
12+log(O/H)(PVT10)\                  & ~7.96$\pm$0.08~~       & ...                & ~7.76$\pm$0.14~~    &      ...               & ~8.35$\pm$0.11~~    &      ...               \\
12+log(O/H)(PM10)\                   &      ...              & ~7.28$\pm$0.17~~    & ~7.64$\pm$0.08~~    & ~7.83$\pm$0.08~~      & ~8.38$\pm$0.08~~     & ~7.87$\pm$0.08~~      \\
\hline
\rule{0pt}{10pt} Value                            & J0810+1837
& J0812+4836          & NGC~2537a           & NGC~2541           &
J0843+4025          & J0911+3135                  \\ \hline
$T_{\rm e}$(OIII)(K)\                & 15,446$\pm$2556~~   & 22,363$\pm$2233~~   & 21,521$\pm$5311~~   & 14,923$\pm$281 ~~  & 17,117$\pm$1317~~   & 17,693$\pm$1918~~    \\
$T_{\rm e}$(OII)(K)\                 & 14,119$\pm$2807~~   & 16,280$\pm$463 ~~   & 16,092$\pm$5459~~   & 13,866$\pm$145 ~~  & 14,707$\pm$1358~~   & 14,831$\pm$1949~~    \\
O$^{+}$/H$^{+}$($\times$10$^5$)\     & 4.270$\pm$2.890~~   & ...                 & 0.765$\pm$1.086~~   & 0.977$\pm$0.073~~  & 1.994$\pm$0.639~~   & 1.861$\pm$0.970~~    \\
O$^{++}$/H$^{+}$($\times$10$^5$)\    & 1.927$\pm$0.800~~   & 0.377$\pm$0.078~~   & 0.766$\pm$0.371~~   & 7.403$\pm$0.383~~  & 1.626$\pm$0.292~~   & 1.195$\pm$0.295~~    \\
O$^{+++}$/H$^{+}$($\times$10$^5$)\    & ...                 & 0.025$\pm$0.026~~   & ...                 & 0.048$\pm$0.023~~  & ...                 & 0.311$\pm$0.352~~    \\
O/H($\times$10$^5$)\                 & 6.197$\pm$2.999~~   & $>$0.402$\pm$0.082~~ & 1.531$\pm$1.148~~   & 8.428$\pm$0.391~~  & 3.621$\pm$0.703~~   & 3.367$\pm$1.073~~    \\
12+log(O/H)\                         & ~7.79$\pm$0.21~~    & $>$6.60$\pm$0.09~~   & ~7.18$\pm$0.33~~    & ~7.93$\pm$0.02~~   & ~7.56$\pm$0.08~~    & ~7.53$\pm$0.14~~     \\
& & & & & &\\
12+log(O/H)(PM10)\                   & ~7.79$\pm$0.14~~    & ~7.23$\pm$0.09~~    & ~8.40$\pm$0.08~~    & ~7.96$\pm$0.08~~   & ~7.63$\pm$0.11~~    & ~7.54$\pm$0.14~~     \\
\hline
Value                             & J0928+2845         &
J0931+2717          & J0940+4459          & KISSB23             &
J0947+4138          & J0951+3842                   \\ \hline
$T_{\rm e}$(OIII)(K)\                & 14,973$\pm$10346~~ & 18,383$\pm$1588~~   & 18,440$\pm$1807~~   & 17,423$\pm$1076~~   & 16,198$\pm$501 ~~   & 14,041$\pm$1733~~     \\
$T_{\rm e}$(OII)(K)\                 & 13,891$\pm$11620~~ & 14,925$\pm$1596~~   & 14,931$\pm$1814~~   & 14,778$\pm$1100~~   & 14,426$\pm$533 ~~   & 13,362$\pm$1083~~     \\
O$^{+}$/H$^{+}$($\times$10$^5$)\     & ...                & 1.557$\pm$0.654~~   & 1.217$\pm$0.548~~   & 1.594$\pm$0.342~~   & 0.702$\pm$0.158~~   & ...                   \\
O$^{++}$/H$^{+}$($\times$10$^5$)\    & 2.527$\pm$4.422~~  & 1.078$\pm$0.233~~   & 1.221$\pm$0.340~~   & 1.597$\pm$0.225~~   & 5.093$\pm$0.383~~   & 5.562$\pm$1.857~~     \\
O$^{+++}$/H$^{+}$($\times$10$^5$)\   & ...                & 0.402$\pm$0.390~~   & ...                 & ...                 & 0.123$\pm$0.038~~   & 0.805$\pm$1.227~~     \\
O/H($\times$10$^5$)\                 & $>$2.527$\pm$4.422~~ & 3.037$\pm$0.797~~ & 2.438$\pm$0.645~~   & 3.192$\pm$0.409~~   & 5.917$\pm$0.416~~   & $>$6.367$\pm$2.226~~     \\
12+log(O/H)\                         &     $>$~7.40        & ~7.48$\pm$0.11~~   & ~7.39$\pm$0.11~~     & ~7.50$\pm$0.06~~    & ~7.77$\pm$0.03~~    & $>$7.80$\pm$0.15~~      \\
& & & & & &\\
12+log(O/H)(PVT10)\                  &      ...           &  ...                &      ...            & ~7.63$\pm$0.14~~    &      ...            &      ...              \\
12+log(O/H)(PM10)\                   & ~7.98$\pm$0.11~~  & ~7.75$\pm$0.14~~    & ~7.49$\pm$0.20~~    & ~7.58$\pm$0.09~~    & ~7.64$\pm$0.09~~    & ~7.92$\pm$0.14~~      \\
\hline
\end{tabular}
\end{table*}

\begin{table*}
\centering{
\caption{{\bf Table B.19} Observed and
corrected relative fluxes in the lines of  galaxies outside the
void (BTA)}
\label{t:Intens_novoid1}
\begin{tabular}{l|c|c|c|c|c|c} \hline
\rule{0pt}{10pt} & \MC{2}{c|}{UGC731} & \MC{2}{c|}{J0839+3140} &
\MC{2}{c}{UGC4787}     \\ \hline \rule{0pt}{10pt} & \MC{2}{c|}{01
10 47.02 +49 36 04.2} & \MC{2}{c|}{08 39 49.05 +31 40 52.4} &
\MC{2}{c}{09 07 35.46 +33 16 42.3}     \\ \hline
$\lambda_{0}$(\AA) Ion                  &
F($\lambda$)/F(H$\beta$)&I($\lambda$)/I(H$\beta$) &
F($\lambda$)/F(H$\beta$)&I($\lambda$)/I(H$\beta$) &
F($\lambda$)/F(H$\beta$)&I($\lambda$)/I(H$\beta$)  \\ \hline
3727\ [O\ {\sc ii}]\                           &  1.722$\pm$ 0.060 &  1.627$\pm$ 0.062             &  1.929$\pm$ 0.766 &  2.170$\pm$ 0.944             &  2.346$\pm$ 0.170 &  2.732$\pm$ 0.206            \\
3868\ [Ne\ {\sc iii}]\                         &  0.342$\pm$ 0.023 &  0.323$\pm$ 0.023             &        ...        &        ...                    &        ...        &        ...                   \\
3967\ [Ne\ {\sc iii}]\ +\ H7\                  &  0.205$\pm$ 0.013 &  0.284$\pm$ 0.022             &        ...        &        ...                    &  0.133$\pm$ 0.015 &  0.209$\pm$ 0.031            \\
4101\ H$\delta$\                               &  0.162$\pm$ 0.009 &  0.239$\pm$ 0.016             &        ...        &        ...                    &  0.227$\pm$ 0.013 &  0.294$\pm$ 0.023            \\
4340\ H$\gamma$\                               &  0.442$\pm$ 0.016 &  0.481$\pm$ 0.019             &        ...        &        ...                    &  0.397$\pm$ 0.017 &  0.459$\pm$ 0.023            \\
\end{tabular}
 }
\end{table*}

\begin{table*}
\centering{
\caption{{\bf Table B.20} Observed and
corrected relative fluxes in the lines of  galaxies outside the
void (BTA) } \label{t:Intens}
\begin{tabular}{l|c|c|c|c|c|c} \hline
\rule{0pt}{10pt} & \MC{2}{c|}{KUG1004+39} & \MC{2}{c|}{UGC5451a} &
\MC{2}{c}{UGC5451b}     \\ \hline \rule{0pt}{10pt} & \MC{2}{c|}{10
07 22.84 +38 58 20.1} & \MC{2}{c|}{10 07 16.93 +47 00 26.7} &
\MC{2}{c}{10 07 17.90 +47 00 20.0}     \\ \hline
$\lambda_{0}$(\AA) Ion                  &
F($\lambda$)/F(H$\beta$)&I($\lambda$)/I(H$\beta$) &
F($\lambda$)/F(H$\beta$)&I($\lambda$)/I(H$\beta$) &
F($\lambda$)/F(H$\beta$)&I($\lambda$)/I(H$\beta$) \\ \hline
3727\ [O\ {\sc ii}]\                           &  1.076$\pm$ 0.031 &  1.097$\pm$ 0.034             &  2.943$\pm$ 0.111 &  2.897$\pm$ 0.125             &  3.823$\pm$ 0.153 &  4.048$\pm$ 0.197          \\
3868\ [Ne\ {\sc iii}]\                         &  0.445$\pm$ 0.012 &  0.452$\pm$ 0.013             &  0.216$\pm$ 0.016 &  0.211$\pm$ 0.017             &  0.248$\pm$ 0.028 &  0.256$\pm$ 0.033          \\
3967\ [Ne\ {\sc iii}]\ +\ H7\                  &  0.267$\pm$ 0.008 &  0.295$\pm$ 0.010             &  0.035$\pm$ 0.005 &  0.118$\pm$ 0.024             &        ...        &        ...                 \\
4101\ H$\delta$\                               &  0.244$\pm$ 0.009 &  0.268$\pm$ 0.011             &  0.209$\pm$ 0.011 &  0.277$\pm$ 0.017             &  0.100$\pm$ 0.012 &  0.231$\pm$ 0.036          \\
4340\ H$\gamma$\                               &  0.449$\pm$ 0.014 &  0.468$\pm$ 0.015             &  0.406$\pm$ 0.018 &  0.461$\pm$ 0.023             &  0.392$\pm$ 0.026 &  0.499$\pm$ 0.040          \\
\end{tabular}
 }
\end{table*}
\begin{table*}
\centering{
\caption{{\bf Table B.21} Observed and
corrected relative fluxes in the lines of  galaxies outside the
void (BTA) }
\label{t:Intens_novoid2}
\begin{tabular}{l|c|c|c|c|c|c} \hline
& \MC{2}{c|}{J1031+2801} & \MC{2}{c|}{NGC3274a} &
\MC{2}{c}{NGC3274b}     \\ \hline \rule{0pt}{10pt} & \MC{2}{c|}{10
31 55.80 +28 01 33.7} & \MC{2}{c|}{10 32 18.87 +27 39 52.7} &
\MC{2}{c}{10 32 19.71 +27 40 00.9}     \\ \hline
$\lambda_{0}$(\AA) Ion                  &
F($\lambda$)/F(H$\beta$)&I($\lambda$)/I(H$\beta$) &
F($\lambda$)/F(H$\beta$)&I($\lambda$)/I(H$\beta$) &
F($\lambda$)/F(H$\beta$)&I($\lambda$)/I(H$\beta$)  \\ \hline
3727\ [O\ {\sc ii}]\                           &  4.064$\pm$ 0.732 &  4.572$\pm$ 1.073             &  2.424$\pm$ 0.064 &  2.428$\pm$ 0.073             &  3.349$\pm$ 0.088 &  4.032$\pm$ 0.116            \\
3868\ [Ne\ {\sc iii}]\                         &        ...        &        ...                    &  0.339$\pm$ 0.016 &  0.337$\pm$ 0.018             &  0.218$\pm$ 0.013 &  0.255$\pm$ 0.015            \\
3967\ [Ne\ {\sc iii}]\ +\ H7\                  &        ...        &        ...                    &  0.153$\pm$ 0.009 &  0.265$\pm$ 0.021             &        ...        &        ...                   \\
4101\ H$\delta$\                               &        ...        &        ...                    &  0.175$\pm$ 0.009 &  0.273$\pm$ 0.018             &  0.173$\pm$ 0.007 &  0.227$\pm$ 0.011            \\
4340\ H$\gamma$\                               &  0.200$\pm$ 0.042 &  0.469$\pm$ 0.150             &  0.391$\pm$ 0.014 &  0.465$\pm$ 0.020             &  0.446$\pm$ 0.013 &  0.502$\pm$ 0.016            \\
\end{tabular}
 }
\end{table*}

\begin{table*}
\centering{
\caption{{\bf Table B.22} Observed and
corrected relative fluxes in the lines of  galaxies outside the
void (BTA)} \label{t:Intens}
\begin{tabular}{l|c|c|c|c} \hline
\rule{0pt}{10pt} & \MC{2}{c|}{UGC5764a} & \MC{2}{c}{UGC6055}     \\
\hline \rule{0pt}{10pt} & \MC{2}{c|}{10 36 43.14 +31 32 50.1} &
\MC{2}{c}{10 58 32.40 +48 46 53.4}     \\ \hline
$\lambda_{0}$(\AA) Ion                  &
F($\lambda$)/F(H$\beta$)&I($\lambda$)/I(H$\beta$) &
F($\lambda$)/F(H$\beta$)&I($\lambda$)/I(H$\beta$)  \\ \hline
3727\ [O\ {\sc ii}]\                           &  0.999$\pm$ 0.049 &  1.080$\pm$ 0.055             &  2.458$\pm$ 0.626 &  2.686$\pm$ 0.688            \\
3868\ [Ne\ {\sc iii}]\                         &  0.402$\pm$ 0.019 &  0.430$\pm$ 0.021             &        ...        &        ...                   \\
3967\ [Ne\ {\sc iii}]\ +\ H7\                  &  0.258$\pm$ 0.015 &  0.278$\pm$ 0.022             &        ...        &        ...                   \\
4101\ H$\delta$\                               &  0.233$\pm$ 0.014 &  0.248$\pm$ 0.021             &        ...        &        ...                   \\
4340\ H$\gamma$\                               &  0.496$\pm$ 0.020 &  0.515$\pm$ 0.025             &  0.487$\pm$ 0.043 &  0.506$\pm$ 0.057            \\
\end{tabular}
 }
\end{table*}

\clearpage

\begin{table*}
\caption{{\bf Table B.23} Oxygen abundance
in the galaxies outside the Lynx-Cancer void (BTA)}
\label{t:Chem_nonvoid}
\begin{tabular}{l|c|c|c|c|c|c} \hline
Value                                & UGC731             &
J0839+3140        & UGC4787           & KUG1004+392       &
UGC5451c          & UGC5451d           \\ \hline
$T_{\rm e}$(OIII)(K)\                & 11,434$\pm$774   & 14,988$\pm$1921 & 14,159$\pm$2169 & 14,070$\pm$602  & 12,713$\pm$1521  & 20,314$\pm$4129   \\
$T_{\rm e}$(OII)(K)\                 & 11,318$\pm$1064  & 13,899$\pm$2156 & 13,434$\pm$2541 & 13,380$\pm$708  & 12,424$\pm$1936  & 14,882$\pm$4170    \\
O$^{+}$/H$^{+}$($\times$10$^5$)\     & 4.084$\pm$1.516  & 2.508$\pm$1.681 & 3.943$\pm$2.132 & 1.568$\pm$0.240 & 4.926$\pm$2.774  & 3.761$\pm$3.282    \\
O$^{++}$/H$^{+}$($\times$10$^5$)\    & 10.820$\pm$2.289 & 3.710$\pm$1.315 & 2.214$\pm$0.894 & 6.796$\pm$0.782 & 4.404$\pm$1.514  & 1.024$\pm$0.425     \\
O/H($\times$10$^5$)\                 & 14.900$\pm$2.745 & 6.218$\pm$2.134 & 6.157$\pm$2.312 & 8.364$\pm$0.818 & 9.330$\pm$3.160  & 4.785$\pm$3.309     \\
12+log(O/H)\                         & ~8.17$\pm$0.08   & ~7.79$\pm$0.15  & ~7.79$\pm$0.16 & ~7.92$\pm$0.04  & ~7.97$\pm$0.15   & ~7.68$\pm$0.30    \\
& & & & & &\\
12+log(O/H)(PVT10)                  & ~7.88$\pm$0.12   &     ~  \ldots         & ~7.92$\pm$0.13  & ~7.76$\pm$0.17  & ~8.23$\pm$0.12   & ~8.18$\pm$0.12     \\
12+log(O/H)(PM10)\                   & ~8.00$\pm$0.10   &      ~  \ldots        & ~7.78$\pm$0.09  & ~\ldots               & ~8.24$\pm$0.08   & ~8.04$\pm$0.08     \\
\hline  
Value                               & J1031+2801        & NGC3274a
& NGC3274b         & UGC5764a            & UGC6055
\\ \hline
$T_{\rm e}$(OIII)(K)\                & 14,244$\pm$1446 & 12,154$\pm$1171  & 15,316$\pm$1129 & 13,734$\pm$925   & 16,324$\pm$1376     \\
$T_{\rm e}$(OII)(K)\                 & 13,486$\pm$1686 & 11,965$\pm$1540  & 14,059$\pm$1248 & 13,164$\pm$1110  & 14,471$\pm$1456     \\
O$^{+}$/H$^{+}$($\times$10$^5$)\     & 5.841$\pm$2.815 & 4.738$\pm$2.273  & 4.489$\pm$1.303 & 1.499$\pm$0.440  & 2.725$\pm$1.117     \\
O$^{++}$/H$^{+}$($\times$10$^5$)\    & 2.802$\pm$0.766 & 6.810$\pm$1.947  & 2.376$\pm$0.437 & 7.440$\pm$1.359  & 1.766$\pm$0.361     \\
O/H($\times$10$^5$)\                 & 8.643$\pm$2.917 & 11.550$\pm$2.993 & 6.864$\pm$1.375 & 8.939$\pm$1.429  & 4.491$\pm$1.174     \\
12+log(O/H)\                         & ~7.94$\pm$0.15  & ~8.06$\pm$0.11   & ~7.84$\pm$0.09  & ~7.95$\pm$0.07   & ~7.65$\pm$0.11      \\
& & & & & &\\
12+log(O/H)(PVT10)\                   & ...               & ~7.99$\pm$0.11   & ~8.06$\pm$0.10  & ~7.85$\pm$0.14   & ...                   \\
12+log(O/H)(PM10)\                   & ...               & ~7.97$\pm$0.10   & ~7.86$\pm$0.09  & ~7.91$\pm$0.15   & ~7.71$\pm$0.13      \\
\hline
\end{tabular}
\end{table*}

\label{lastpage}

\end{document}